\documentclass[12pt]{article}
\usepackage{epsf, cite, amssymb}
\usepackage{epsfig}

\makeatletter
\renewcommand\section{\@startsection {section}{1}{\z@}%
                                   {-3.5ex \@plus -1ex \@minus -.2ex}%nn
                                   {2.3ex \@plus.2ex}%
                                   {\normalfont\large\bfseries}}
\renewcommand\subsection{\@startsection{subsection}{2}{\z@}%
                                     {-3.25ex\@plus -1ex \@minus -.2ex}%
                                     {1.5ex \@plus .2ex}%
                                     {\normalfont\bfseries}}
\makeatother
\def\IZ{\relax\ifmmode\mathchoice
{\hbox{\cmss Z\kern-.4em Z}}{\hbox{\cmss Z\kern-.4em Z}}
{\lower.9pt\hbox{\cmsss Z\kern-.4em Z}} {\lower1.2pt\hbox{\cmsss
Z\kern-.4em Z}}\else{\cmss Z\kern-.4em Z}\fi}
\def\IR{\relax{\rm I\kern-.18em R}}

\def\one{{\hbox{ 1\kern-.8mm l}}}

\newlength{\bredde}
\def\slash#1{\settowidth{\bredde}{$#1$}\ifmmode\,\raisebox{.15ex}{/}
\hspace*{-\bredde} #1\else$\,\raisebox{.15ex}{/}\hspace*{-\bredde}
#1$\fi}

\newcommand {\Cbar}
    {\mathord{\setlength{\unitlength}{1em}
     \begin{picture}(0.6,0.7)(-0.1,0)
        \put(-0.1,0){\rm C}
        \thicklines
        \put(0.2,0.05){\line(0,1){0.55}}
     \end {picture}}}
\newsavebox{\zzzbar}
\sbox{\zzzbar}
  {\setlength{\unitlength}{0.9em}
  \begin{picture}(0.6,0.7)
  \thinlines
  \put(0,0){\line(1,0){0.6}}
  \put(0,0.75){\line(1,0){0.575}}
  \multiput(0,0)(0.0125,0.025){30}{\rule{0.3pt}{0.3pt}}
  \multiput(0.2,0)(0.0125,0.025){30}{\rule{0.3pt}{0.3pt}}
  \put(0,0.75){\line(0,-1){0.15}}
  \put(0.015,0.75){\line(0,-1){0.1}}
  \put(0.03,0.75){\line(0,-1){0.075}}
  \put(0.045,0.75){\line(0,-1){0.05}}
  \put(0.05,0.75){\line(0,-1){0.025}}
  \put(0.6,0){\line(0,1){0.15}}
  \put(0.585,0){\line(0,1){0.1}}
  \put(0.57,0){\line(0,1){0.075}}
  \put(0.555,0){\line(0,1){0.05}}
  \put(0.55,0){\line(0,1){0.025}}
  \end{picture}}
\newcommand{\Zbar}{\mathord{\!{\usebox{\zzzbar}}}}

\newcommand{\ena}{\end{eqnarray}}
\newcommand{\beqa}{\begin{eqnarray}}
\newcommand{\eeqa}{\end{eqnarray}}
\newcommand{\bea}{\begin{eqnarray}}
\newcommand{\eea}{\end{eqnarray}}

\newcommand{\eq}[1]{(\ref{#1})}
\newcommand{\fig}[1]{Figure~\ref{#1}}

\newcommand{\be}{\begin{equation}}
\newcommand{\ee}{\end{equation}}

\begin{document}

\begin{titlepage}
\begin{flushright}
hep-th/0605199
\end{flushright}
\vfill
\begin{center}
{\LARGE\bf Big Bang Models in String Theory%
%\footnote{Based on lectures given at RTN Winter School on Strings, Supergravity and Gauge Theories, CERN, January 16 - 20, 2006}
}    \\
\vskip 15mm
{\large Ben Craps}
%\footnote{email address: Ben.Craps@vub.ac.be} \\
\vskip 10mm
{\em Theoretische Natuurkunde, Vrije Universiteit Brussel and\\
The International Solvay Institutes\\ Pleinlaan 2, B-1050
Brussels, Belgium}
\vskip 5mm
{\tt Ben.Craps@vub.ac.be}
\end{center}
\vfill

\begin{center}
{\bf ABSTRACT}
\end{center}
%\begin{quote}
These proceedings are based on lectures delivered at the ``RTN Winter School on Strings, Supergravity and Gauge Theories'', CERN, January 16 - January 20, 2006. The school was mainly aimed at Ph.D.\ students and young postdocs. The lectures start with a brief introduction to spacetime singularities and the string theory resolution of certain static singularities. Then they discuss attempts to resolve cosmological singularities in string theory, mainly focusing on two specific examples: the Milne orbifold and the matrix big bang. 
%\vskip 7mm
\vfill
%\end{quote}
%\begin{flushleft}
%PACS 11.25.-w, 04.65.+e
%\end{flushleft}
\end{titlepage}
%%%%%%%%%%%%%%%%%%%%%%%%%%%%%%%%%%%%%%%%%%%%%%%%%%%%%%%%%%%%%%%%%%%%%%%%%%%%%%%%%%%%%%%%%%%%%%%%%%%%%%%%%%%%%%%%%%%%%%%%%%%%%%%%%%%%%%%%%%%%%%%%%%%%%%%%%%%%%%%%%%%%%%%%%%%%%%%%%%%%%%%%%%%%%%%%%%%%%%
\section{Lecture I: Introduction and outline}

Spacetime singularities are a prediction of general relativity. Examples include singularities inside black holes and the big bang singularity at the beginning of a Friedmann-Lema\^itre-Robertson-Walker (FLRW) universe. However, general relativity breaks down at spacetime singularities, which is one of the main motivations for pursuing a consistent quantum theory of gravity. The best developed candidate is string theory, and the purpose of these lectures is to explore what string theory has to say about spacetime singularities.
\begin{figure}
\begin{center}
\epsfig{file=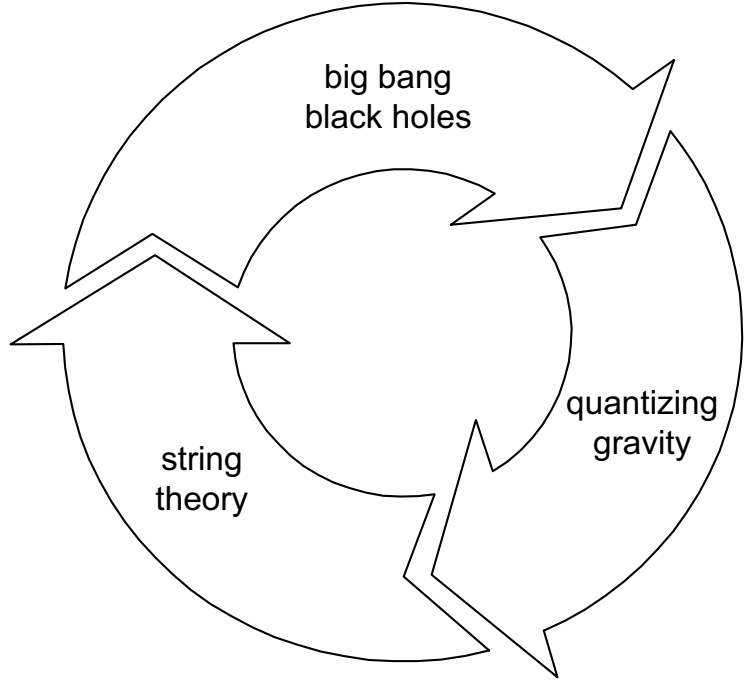, width=10cm}
\end{center}
\caption{Can string theory make sense of spacetime singularities?}\label{figSingularities}
\end{figure}

Rather than attempt to give a broad overview of this subject, I will focus on a number of specific models and techniques. Complementary reviews include \cite{Gasperini:2002bn, Quevedo:2002xw, Giveon:2003gb, Cornalba:2003kd, Durin:2005ix}.

In this first lecture, I will first briefly introduce spacetime singularities in the context of general relativity, and argue that one should go beyond general relativity to try to really understand them. Next, I will indicate how certain static singularities are resolved in string theory, focusing in particular on orbifold and ALE singularities. Finally, we will have a first look at ``cosmological'' (spacelike or lightlike) singularities in string theory, which will be the focus of the remainder of these lectures. 
%%%%%%%%%%%%%%%%%%%%%%%%%%%%%%%%%%
\subsection{General relativity and spacetime singularities}  

In general relativity, freely falling particles follow geodesics in a curved spacetime. What is important for these lectures, is that geodesics don't always go on forever: they can end in a ``singularity'', which can intuitively be thought of as a ``place of infinite curvature''. 
\begin{figure}
\begin{center}
\epsfig{file=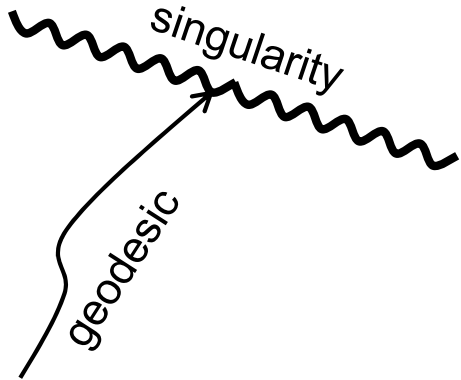, width=7cm}
\end{center}
\caption{A geodesic ending in a singularity.}\label{figGeodesic}
\end{figure}
General relativity doesn't tell us what happens when a particle hits a singularity.

A well-known example of a singularity is the big bang singularity of a cosmological spacetime. An FLRW spacetime (with flat spatial sections) is described by a metric
\be
ds^2=-dt^2+a^2(t)\left[dx^2+dy^2+dz^2\right],
\ee
where for a radiation or matter dominated universe the scale factor $a(t)$ goes to zero at some finite time $t=0$, where the spacetime is infinitely curved. General relativity does not tell us what happens at $t=0$.

Another famous example are the singularities inside black holes. General relativity predicts the existence of black holes, but breaks down at their singularities.

Clearly, a more fundamental theory is needed if we want to make sense of spacetime singularities, in particular if we want to know the laws of physics that govern the dynamics of particles when they hit a spacetime singularity. Since quantum effects are expected to be important when the spacetime curvature reaches Planckian values, it is natural to look for a quantum theory of gravity. However, it is well-known that quantizing general relativity as a fundamental theory is problematic since the theory is not renormalizable: it contains divergences that cannot be cured by redefining a finite number of parameters.

String theory manages to ``smear out'' the gravitational interactions, leading to a consistent quantization of gravity. Pictorially, Feynman diagrams are replaced by stringy Feynman diagrams, where lines are replaced by tubes. This delocalizes the interaction vertices. From the string theory point of view, general relativity should be viewed as a low-energy effective theory.

Let us pauze a moment to discuss (non-)renormalizability and effective field theory. A well-known example of a renormalizable field theory is quantum electrodynamics (QED). Divergences in QED loop diagrams can be absorbed in the redefinition of a finite number of physical parameters, such as the mass and charge of the electron. Once these parameters have been measured, the theory makes finite predictions for all other physical observables. Such a theory is called ``renormalizable''.

An example of a non-renormalizable field theory is Fermi's four fermion model of the weak interactions, whose coupling constant $G_F$ has dimension length squared. As a consequence, higher loop diagrams lead to divergences involving higher powers of $G_F\Lambda^2$, where $\Lambda$ is an ultraviolet cutoff. Renormalizing away all these divergences would require introducing an infinite number of terms with undetermined coefficients. Therefore the theory is only useful in a regime where one expects almost all of these terms to be negligible, namely for low enough energies. The four fermion model is a low-energy effective theory, breaking down at energies of order $(G_F)^{-1/2}\sim 300$ GeV.

For the case of Fermi's theory, we know how to do better: in the electroweak theory, Fermi's four-fermion contact interaction is replaced by a Feynman diagram with two three-point vertices and a W-boson propagator connecting the two vertices. 
\begin{figure}
\begin{center}
\epsfig{file=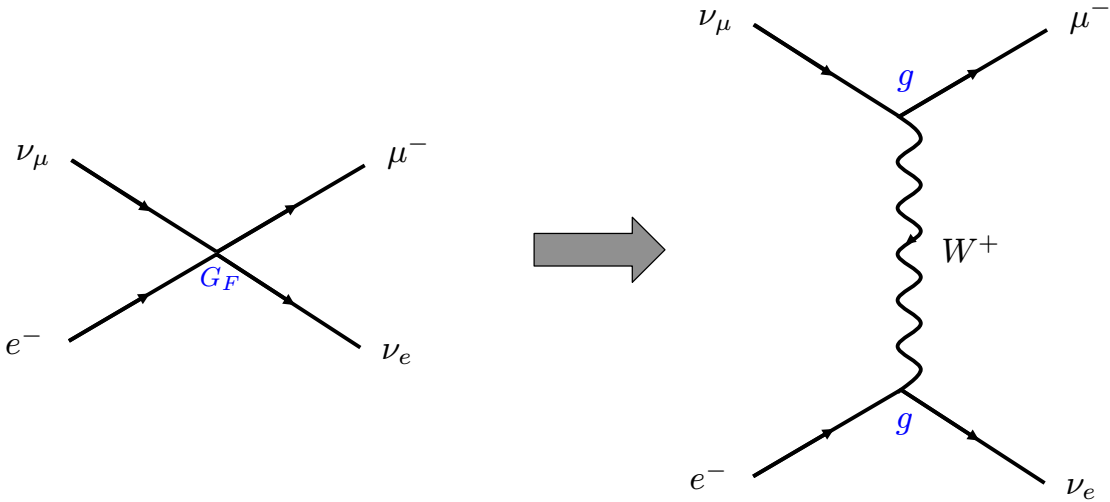, width=16.3cm}
\end{center}
\caption{Four fermion interaction replaced by a W-boson exchange diagram.}\label{figweak}
\end{figure}
At sufficiently low energies, Fermi's model is a good approximation, as can be seen by expanding the W-boson propagator as
\be
{1 \over k^2+M_W^2}={1\over M_W^2}-{k^2\over M_W^4}+\ldots,
\ee 
where $k^2$, the square of the four-momentum of the W-boson, is supposed to be small compared to its mass squared. The first term in this expansion can be identified with Fermi's constant,
\be
G_F\sim {g^2\over M_W^2}.
\ee
We see that the W-boson can be ignored (i.e.\ its effects can be replaced by an effective contact interaction) as long as it is much heavier than the energy scales of interest. The fact that heavy particles do not have to be included explicitly in a theory is true more generally.

Let us say a few words about the usefulness and limitations of effective field theories (see \cite{Burgess:2003jk} for a nice discussion in the context of quantum gravity). As mentioned before, general relativity (or supergravity) should be thought of as a low-energy effective theory. The underlying fundamental theory (e.g.\ string theory) contains additional ``heavy'' degrees of freedom. The effective theory arises upon integrating out those heavy degrees of freedom:
\be
e^{iS_{\rm eff}(l)}=\int Dh\, e^{iS(l,h)},
\ee
where ``$l$'' denotes the ``light'' (super)gravity fields, while ``$h$'' denotes the ``heavy'' additional fields. The effective action can be expanded in small parameters like $E^2/M^2$, where $E$ is the (low) energy scale of interest and $M$ is the mass of a heavy mode. Meaningful quantum gravity effects can be computed in this expansion. The effective theory is not useful when an expansion parameter is not small. For instance, the mass of a ``heavy'' mode may depend on the light modes and may actually be small in certain regimes. In extreme such cases, when the mass of a "heavy" mode goes to zero, the effective theory can become singular (it can contain divergent couplings). The resolution typically consists in not integrating out a ``heavy'' mode when it becomes light, but including it in the low-energy description. The additional (e.g.\ stringy) degrees of freedom ``resolve'' the singularity of the original gravity theory.

To summarize, we have seen that general relativity is an effective theory. It breaks down near spacetime singularities (where the curvature diverges). A possible explanation for the breakdown is that additional degrees of freedom from the underlying fundamental theory become light and should not be integrated out. We have a promising candidate for the underlying theory, namely string theory.

The question then becomes: do stringy degrees of freedom resolve spacetime singularities? Stringy degrees of freedom come in two varieties: perturbative (the strings themselves) and non-perturbative (such as branes). We will see both types of degrees of freedom at work in resolving spacetime singularities.                        
 
%%%%%%%%%%%%%%%%%%%%%%%%%%%%%%%%%%
\subsection{Static singularities in string theory}

The simplest examples of singularities in string theory are orbifold \cite{Dixon:1985jw, Dixon:1986jc} singularities. In the simplest context, orbifolds can be thought of as manifolds quotiented by discrete groups, i.e.\ manifolds with discrete identifications. Consider, for example, a two-dimensional plane and identify points that are related by a rotation over $360/n$ degrees with integer $n\ge2$. The resulting space is a cone, the tip of which is a singular point with infinite curvature. In a space containing such a cone, general relativity is unable to say what happens when a particle hits the tip of the cone. The situation is completely different in string theory: it turns out that string perturbation theory is perfectly smooth on orbifold spacetimes. The reason is that perturbative string theory automatically takes into account the so-called ``twisted closed strings'', which are open strings on the covering space that become closed after the identification (strings winding around the tip of the cone). Twisted closed strings provide light degrees of freedom that were not taken into account in general relativity. These additional light degrees of freedom ``resolve'' the singularity within perturbative string theory, meaning that one obtains sensible, finite answers to all physical questions.

Particularly interesting examples of orbifolds \cite{Douglas:1996sw} are of the type $\Cbar^2/\Zbar_n$, where the $\Zbar_n$ identification is
\be
(z^1,z^2)\sim (e^{2\pi i/n}z^1,e^{-2\pi i/n}z^2).
\ee
The orbifold has a singularity at the fixed point $(z^1,z^2)=(0,0)$; mathematically, it is known as an $A_{n-1}$ singularity, a special case of an ALE singularity. Once again, perturbative string theory turns out to be completely smooth due to the twisted closed strings. It is well-known that geometrically, the $A_{n-1}$ singularity can be resolved into $n-1$ intersecting two-spheres. Before saying more about this singularity, I have to briefly introduce the concept of D-branes.

D-branes are extended objects in string theory with the defining property that open strings can end on them. The oscillation modes of the open strings are the degrees of freedom of the D-branes. They include scalar fields $X^i$ describing the position and profile of the brane in its transverse dimensions. 
\begin{figure}
\begin{center}
\epsfig{file=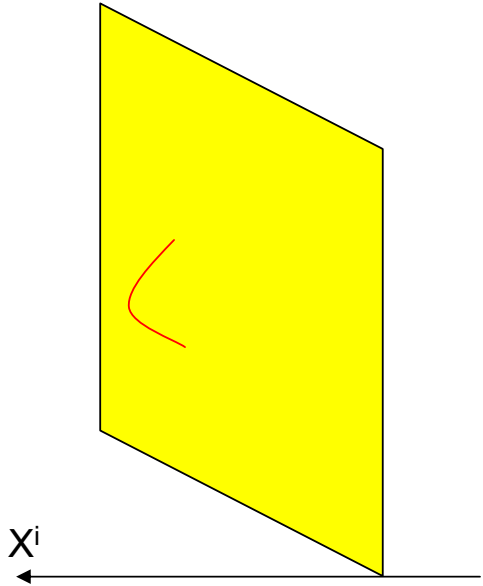, width=7cm}
\end{center}
\caption{Scalar fields from open strings describe the position and profile of a D-brane in its transverse dimensions $X^i$.}\label{figDbrane}
\end{figure}
The tension of a D-brane turns out to be proportional to $1/g_s$ (with $g_s$ the string coupling constant), which is very large at weak coupling. One of the uses of D-branes is that they can wrap cycles in the extra dimensions of string theory, thus giving rise to charged particles in the lower-dimensional effective theory. The mass of such a particle equals the product of the D-brane tension and the volume of the wrapped cycle. At weak coupling, they are thus very heavy unless the brane wraps a very small cycle.

For multiple D-branes, additional structure appears. For two D-branes, say, one expects at least two sets of scalar fields, $X^i_{11}$ and $X^i_{22}$, labelling the positions and profiles of both branes. These arise from open strings beginning and ending on the same brane. 
\begin{figure}
\begin{center}
\epsfig{file=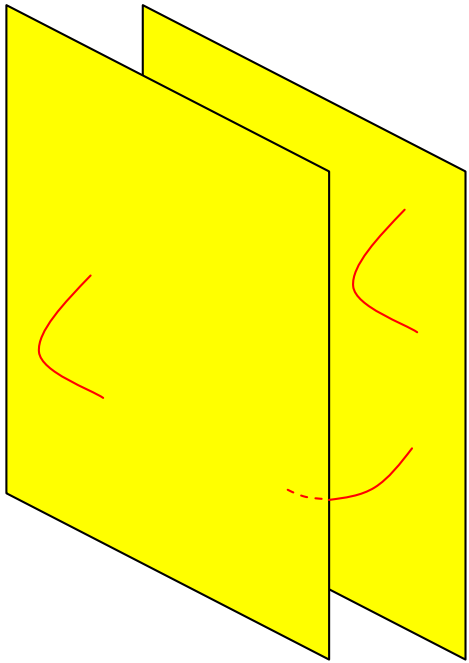, width=7cm}
\end{center}
\caption{Open strings stretching between two different D-branes give rise to off-diagonal matrix degrees of freedom.}\label{figMultiple}
\end{figure}
However, one finds more fields, corresponding to open strings stretching between the two branes, $X^i_{12}$ and $X^i_{21}$. As I have anticipated in the notation, all these fields can be naturally combined in a $2\times2$ matrix
\be
X^i=\left( \begin{array}{cc} X^i_{11} & X^i_{12}  \\ X^i_{21} & X^i_{22}  \end{array} \right) .
\ee
It turns out that the effective action describing multiple D-branes contains a potential term
\be
V\sim{\rm Tr}[X^i,X^j]^2.
\ee
This implies that the off-diagonal modes (stretched strings) are very massive when the branes are well-separated. Then only the diagonal modes (brane positions/profiles) are light. On the other hand, when the branes are close to each other, all the matrix degrees of freedom are light. This is a first indication that the notion of spacetime becomes ``fuzzy'' at short distances: on top of spacetime positions, there are additional light degrees of freedom that have to be taken into account to get a non-singular description.

Going back to orbifold singularities, it turns out that D-branes are useful probes of the structure of the singularities. A systematic framework to study D-branes is boundary conformal field theory, i.e.\ conformal field theory on two-dimensional manifolds with boundaries \cite{Recknagel:1997sb, Fuchs:1997fu}. For the orbifolds we are considering, one finds that there exist two types of D-branes. The first type are ``bulk'' D-branes. On the covering space, they correspond to a $\Zbar_n$ symmetric configuration of $n$ D-branes. Bulk branes have the property that they can move anywhere in the orbifold: the images move in such a way that the configuration remains symmetric under the $\Zbar_n$ orbifold group. The second type are ``fractional'' D-branes. On the covering space, they correspond to just a single D-brane placed at the fixed point, which is a symmetric configuration by itself. Fractional branes are stuck at the orbifold singularity: to move away from it, they would need $n-1$ companions to keep the configuration symmetric, but those companions aren't there. The fact that fractional branes are stuck at the singularity makes them ideal probes of the structure of the singularity. We have seen before that an $A_{n-1}$ singularity can be viewed as a limit of the resolved $A_{n-1}$ singularity, where the $n-1$ two-spheres collapse into a single point. A precise correspondence has been established between the fractional D-branes of the orbifold and D-branes wrapping the two-spheres of the resolved orbifold \cite{Douglas:1996sw, Polchinski:1996ry, Diaconescu:1997br, Billo:2000yb}. 
\begin{figure}
\begin{center}
\epsfig{file=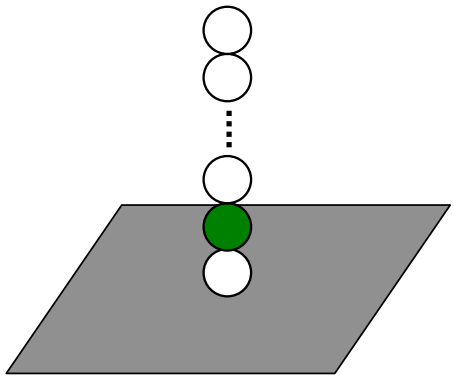, width=10cm}
\end{center}
\caption{Fractional branes as wrapped branes.}\label{figALE}
\end{figure}
Using this correspondence, topological information of the resolved orbifold, such as the intersection numbers of the two-spheres, can be easily computed as a certain index in conformal field theory \cite{Douglas:1999hq}.

At this point, there may seem to be a contradiction between two things I have said. On the one hand, I have said that orbifold singularities are completely resolved within perturbative string theory, because twisted closed strings are taken into account in that framework. On the other hand, I have mentioned that $\Cbar^2/\Zbar_n$ orbifolds contain non-perturbative objects, namely fractional D-branes, that correspond to D-branes wrapping cycles that vanish in the orbifold limit. But wouldn't D-branes wrapping vanishing cycles be expected to be massless, and wouldn't the presence of massless non-perturbative degrees of freedom imply that the singularity cannot be resolved in perturbative string theory?  

The resolutions is that in string theory cycles are characterized not only by their size, but also by fluxes through them. It turns out that the orbifold conformal field theory corresponds to a non-vanishing flux of the NS-NS two-form potential $B$ through the vanishing cycles \cite{Aspinwall:1995zi}. This B-flux gives the wrapped D2-branes a mass. One way to see that the B-flux should give rise to a mass is as follows. Consider the Wess-Zumino term in the effective action for a D2-brane,
\be
S^{D2}_{WZ}\sim\int C_{3}+B\wedge C_{1}+\ldots
\ee
This shows that B-flux induces D0-brane charge on a wrapped D2-brane. Because of the BPS-bound, which says that in certain units the mass is bounded from below by the absolute value of the charge, this implies a non-vanishing mass.

This discussion raises another question: what happens if the B-flux is turned off? This can be done by a marginal deformation in conformal field theory and corresponds to going to a different point in moduli space. In the absence of B-flux, the D-branes wrapping vanishing cycles are now massless. Since these are non-perturbative degrees of freedom, string perturbation theory is expected to 
break down. Indeed, it turns out that string perturbation theory is now singular! The conclusion is that wrapped D-branes are light and need to be included in the description. This is very similar to what happens for the conifold \cite{Strominger:1995cz}. The quantitative study of these singularities in string theory is much more challenging than for orbifolds. Techniques include (double scaled) little string theory \cite{Ooguri:1995wj, Berkooz:1997cq, Seiberg:1997zk, Giveon:1999px}.

We conclude that including wrapped D-branes in the description resolves the $A_{n-1}$ singularity without B-flux in non-perturbative string theory.

To summarize, at (static) orbifold singularities, perturbative string modes (twisted closed strings) are light. They resolve the singularity in perturbative string theory. Without the B-flux, there are also light non-perturbative modes (wrapped D-branes). The singularity is then resolved in non-perturbative string theory. String theory thus successfully resolves (important classes of) static singularities. In what follows, we will explore to what extent this success can be extended to cosmological singularities.

%%%%%%%%%%%%%%%%%%%%%%%%%%%%%%%%%%
\subsection{A first look at cosmological singularities in string theory}

In recent years, a lot of work has been done on the interface between string theory and cosmology. One important issue is that of cosmological singularities. The concept of a big bang raises a number of deep questions: Is the big bang the beginning of time? Or was there something before the big bang? If so, how does one go through the singularity? The study of these questions in string theory was pioneered in the ``pre-big-bang scenario'' \cite{Gasperini:1992em} (see \cite{Gasperini:2002bn} for a review). Recently, these issues also came up in the string-inspired ekpyrotic and cyclic universe scenarios, where the density perturbations that are now observed as CMB anisotropies were created before the big bang \cite{Khoury:2001bz, Khoury:2001zk, Steinhardt:2001st}. Therefore, it is crucial to know how these perturbations pass through a big crunch/big bang singularity. But even in the more mainstream inflationary universe, the details of the initial state can be important (for instance to understand how inflation started in the first place). It is hoped that string theory will lead to a detailed understanding of the big bang, which in turn may lead to observable signatures of string theory. 

So does string theory resolve spacelike (and lightlike) singularities? The best understood examples of static singularities are orbifolds. So let's now consider time-dependent orbifolds. Techniques from coset conformal field theory have proven useful too, though we will not discuss them in detail in these lectures. 

One typically finds that the propagation of free untwisted fluctuations through a big crunch/big bang  singularity is under control (since global untwisted wavefunctions on the orbifold are simply wavefunctions on the covering space satisfying an invariance condition). We will briefly sketch one example momentarily. However, it has turned out to be problematic to include interactions. Models with spacelike or lightlike singularities tend to suffer from large gravitational backreaction. Efforts to get around those problems are still ongoing. This will be discussed in the second lecture. 

Time-dependent orbifolds often (but not always) have closed timelike curves. This has led to lots of studies of the role of closed causal curves in string theory, in particular to mechanisms that can excise regions with closed timelike curves. We will not discuss these developments in these lectures. Instead, we will naively attempt to compute S-matrix elements relating fluctuations in the far past to fluctuations in the far future, even if there are intermediate regions with closed timelike curves. The main motivation is that, at least in the backgrounds where string theory is best understood, the fundamental observables always have to do with asymptotic regions of spacetime, such as S-matrix elements in asymptotically Minkowski space and boundary correlation functions in asymptotically anti-de Sitter space. If we find a sensible S-matrix, we will go back to interpretational issues related to closed timelike curves. Otherwise, we will look for other techniques or better models.

In particular, if string perturbation theory doesn't give a satisfactory resolution, we will have to rely on the non-perturbative formulations of string theory we have at our disposal. That will be the subject of roughly the second half of these lectures.

\begin{figure}
\begin{center}
\epsfig{file=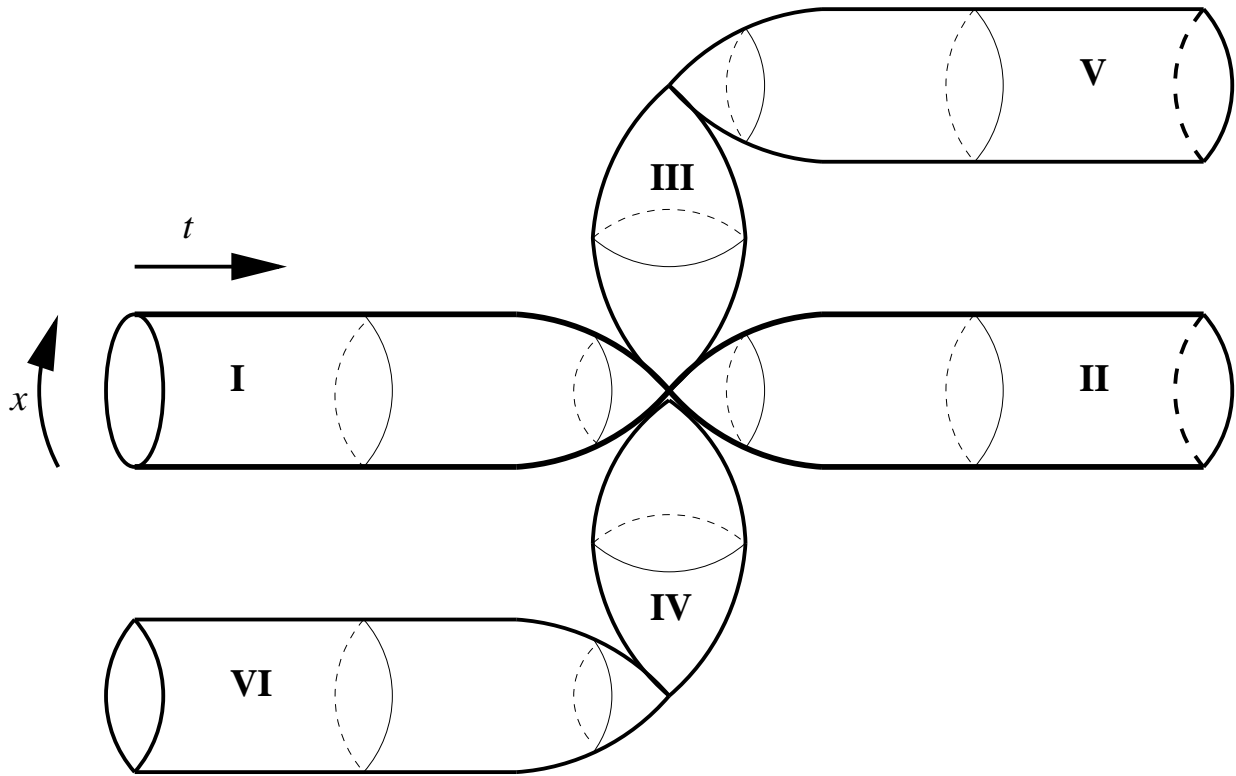, width=16.3cm}
\end{center}
\caption{A time-dependent orbifold \cite{Craps:2002ii}.}\label{figCKR}
\end{figure}

As a first example of a time-dependent orbifold, consider \fig{figCKR} \cite{Craps:2002ii}. 
It shows two dimensions of a higher-dimensional spacetime, which is an exact solution of classical string theory. Region I contains a circle that shrinks to zero size at some time $t=0$. In region II, the circle expands again. The singularity at $t=0$ is a spacelike orbifold singularity. The ``intermediate'' regions III and IV contain closed timelike curves and are connected to additional ``asymptotic'' regions V and VI. One could ask why we bother to study such a complicated spacetime, instead of just regions I and II. It turns out that the rules of string theory force us to include all six regions: no consistent classical solution is known that consists of only regions I and II. Using orbifold and conformal field theory technology, it is found that free strings are able to propagate through the singularity and that particles are created at the singularity \cite{Craps:2002ii} (see \cite{Elitzur:2002rt} for a similar analysis of a closely related model).

Imagine an observer going from region I to region II. The circle in the $x$ direction is part of the internal space. 
\begin{figure}
\begin{center}
\epsfig{file=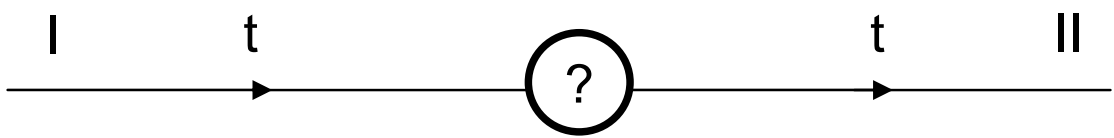, width=16.3cm}
\end{center}
\caption{Four-dimensional spacetime corresponding to the time-dependent orbifold of \fig{figCKR} \cite{Craps:2003ai}.}\label{figQuestion}
\end{figure}
The time coordinate $t$ and three infinite space dimensions make up the 
conventional four-dimensional spacetime. The four-dimensional spacetime metric describes an FLRW cosmology, but the scale factor of the metric vanishes at $t=0$: the four-dimensional description breaks down. Indeed, strings winding around the $x$ circle become light near $t=0$, so the internal dimensions cannot be ignored. From the higher-dimensional string theory description, we know how free fluctuations pass through the singularity, so we can ``resolve'' the question mark in \fig{figQuestion} and compute the density perturbations before and after the crunch. Time evolution from I to II fails to be unitary, because information can leak in and out from the additional regions predicted by string theory \cite{Craps:2003ai}.
 
Locally near the singularity between regions I and II, the previous spacetime is well-approximated by the Milne orbifold, which is Minkowski space with a discrete boost identification. Start with the Minkowski metric
\be
ds_{10}^2=-2dX^+dX^-+(dX^i)^2
\ee
and identify points related by a discrete boost transformation,
\be
X^\pm\sim e^{\pm2\pi}X^\pm.
\ee
The resulting spacetime has a singularity at the fixed point $X^\pm=0$. The Milne orbifold consists of four cones touching each other at their common tip (where the space is actually non-Hausdorff): a ``future'' cone, a ``past'' cone and two ``whisker'' regions with closed timelike curves. 
\begin{figure}
\begin{center}
\epsfig{file=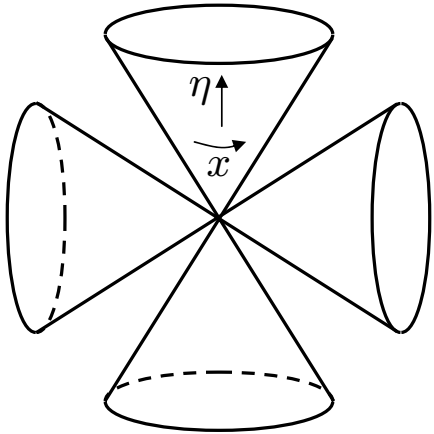, width=7cm}
\end{center}
\caption{The Milne orbifold.}\label{figMilne}
\end{figure}
To make the geometry more transparent, let us introduce new coordinates covering the future cone $X^\pm>0$:
\be
X^\pm={1\over\sqrt2}e^{\eta\pm x},
\ee
in terms of which the metric reads
\be
ds_{10}^2=e^{2\eta}(-d\eta^2+dx^2)+(dX^i)^2.
\ee
In these coordinates, the identification is simply
\be
x\sim x+2\pi
\ee
and the singularity is at $\eta=-\infty$. The circle parametrized by $x$ will be called the ``Milne circle''.
  
Natural objects to compute in string theory are S-matrix elements: send in a number of particles into the past cone, and see what comes out in the future cone. As we have mentioned before, our attitude is to try and apply the usual rules of orbifold conformal field theory (which worked very well for static orbifold singularities) without worrying about apparent pathologies like closed timelike curves. If there are true pathologies, they will presumably show up at some point in the computations of S-matrix elements. The orbifold procedure immediately tells us how free particles propagate through the singularity: globally defined wavefunctions are simply given by orbifold invariant superpositions of the well-known flat space wavefunctions. The question is what happens when interactions are taken into account.

One finds \cite{Berkooz:2002je} that at string tree level, the two and three point functions in the Milne orbifold give sensible results, but that the four point function (corresponding to two to two scattering) exhibits unusual divergences. We will see this in detail in the next lecture. These divergences show that the singularity is not resolved in perturbative string theory. The cause of the divergences can be easily understood. As the circle shrinks, the perturbations undergo an infinite blueshift and create a large gravitational field. Tree-level gravitational interaction with the second perturbation causes the divergence \cite{Berkooz:2002je}. One might be tempted to think that perhaps the failure of string perturbation theory is related to the presence of closed timelike curves in the whisker regions, or to the fact that we are considering a non-supersymmetric orbifold. However, in earlier work \cite{Liu:2002ft, Liu:2002kb}, it was found that a similar but supersymmetric, lightlike orbifold singularity (of the parabolic orbifold \cite{Horowitz:1991ap, Simon:2002ma}) is not resolved in string perturbation theory either. A non-perturbative manifestation of gravitational backreaction was pointed out in \cite{Horowitz:2002mw}: when a single particle is added to the spacetime, its backreaction will lead to the creation of a large black hole, thus turning the apparently simple orbifold singularity in a more generic spacelike singularity. See \cite{Lawrence:2002aj} for a related analysis. The precise relation between this non-perturbative effect and the tree-level divergences is not entirely clear (see also \cite{Cornalba:2003ze}). It is still unknown whether the singularity of the Milne orbifold is resolved non-perturbatively in string theory.

To summarize, the study of cosmological singularities in string theory is motivated by the following questions: what are the laws of physics near big bang and black hole singularities? Does it make sense to ask what happened before the big bang? Time-dependent orbifolds allow us to compute how free fluctuations propagate through a big crunch/big bang singularity. However, interactions lead to unusual divergences and the breakdown of string perturbation theory, as we will discuss in more detail in the next lecture.  

The outline of the remaining three lectures is as follows. In Lecture II, we will outline the computation of the divergences that invalidate string perturbation theory in the Milne orbifold. We will also discuss efforts to make progress, by including a condensate of cosmologically produced winding strings, by modifying the model so that it includes closed string tachyons and by probing the spacetime with D-instantons. Finally, we will sketch a use of the AdS/CFT-correspondence to study a black hole singularity. In Lectures III and IV, we will introduce the lightlike linear dilaton model, which has a lightlike singularity that appears to admit a matrix theory description. After an introduction to matrix theory in flat spacetime, we will derive the matrix theory description of the lightlike linear dilaton background, the ``matrix big bang''. Finally, we will discuss some aspects of the dynamics of the model and list a number of open questions.

%%%%%%%%%%%%%%%%%%%%%%%%%%%%%%%%%%%%%%%%%%%%%%%%%%%%%%%%%%%%%%%%%%%%%%%%%%%%%%%%%%%%%%%%%%%%%%%%%%%%%%%%%%%%%%%%%%%%%%%%%%%%%%%%%%%%%%%%%%%%%%%%%%%%%%%%%%%%%%%%%%%%%%%%%%%%%%%%%%%%%%%%%%%%%%%%%%%%%
\section{Lecture II: Time-dependent orbifolds and AdS/CFT}

In Lecture I, we argued that general relativity is a low-energy effective theory and arises from a fundamental theory (e.g.\ string theory) by integrating out heavy modes. It breaks down near spacetime singularities. A possible explanation for the breakdown is that additional degrees of freedom from the fundamental theory become light and should not have been integrated out. When all light degrees of freedom are present in the description, the description should be smooth. For orbifold singularities, the additional light degrees of freedom are twisted closed strings. String perturbation theory takes them into account automatically; indeed, perturbative string amplitudes are finite. Orbifold singularities are resolved in perturbative string theory. In other examples, such as ALE singularities without B-flux, there are light non-perturbative modes, such as D-branes wrapping vanishing cycles. Indeed, in those examples, perturbative string amplitudes are singular. One needs non-perturbative string theory to resolve this type of singularities. The question is whether string theory also resolves non-static singularities, which are of interest for cosmology. Time-dependent orbifolds allow us to compute how free fluctuations propagate through a big crunch/big bang singularity. However, interactions lead to unusual divergences and the breakdown of string perturbation theory.

Now we will discuss the tree-level divergences in the Milne orbifold in more detail. We will go through the computation of tree-level two to two scattering amplitudes in the Milne orbifold and find unusual divergences. These divergences have been associated with large tree-level gravitational backreaction. A simple observation is that since the tree-level result is supposed to be the first term in a series expansion in the string coupling constant, it being infinite implies that string perturbation theory breaks down. We will briefly review recent efforts to save string perturbation theory by including effects of twisted closed strings that could lead to the singularity (and the associated divergences) being avoided. At the end of this lecture, we briefly sketch a first non-perturbative approach to the study of cosmological singularities, based on the AdS/CFT correspondence.

%%%%%%%%%%%%%%%%%%%%%%%%%%%%%%%
\subsection{Tree-level amplitudes of untwisted states in the Milne orbifold}

Scalar wavefunctions on the orbifold are scalar wavefunctions on Minkowski space invariant under the boost identification $X^\pm\sim e^{\pm2\pi}X^\pm$ \cite{Nekrasov:2002kf}:
\be
\psi_{m,l}(X^\pm,\vec X)=\int_{-\infty}^\infty dw\exp\left(ip^+X^-e^{-w} +ip^-X^+e^w+i\vec p\cdot\vec X+ilw\right)
\ee                               
with $l\in\Zbar$ and $2p^+p^-=m^2$ being the effective two-dimensional mass squared (we will choose $p^+=p^-=m/\sqrt2$):
\be
m^2={\rm mass}^2+\vec p\,^2.
\ee
In the future cone $X^\pm>0$, the wavefunctions can be written as
\be
\psi_{m,l}\sim e^{-ilx}H^{(1)}_{-il}(me^\eta).
\ee
Note that $l$ is the momentum along the Milne circle. Near the singularity $\eta\to-\infty$, the wavefunctions behave like
\bea
\psi_{m,l}&=& A_{m,l}e^{-il(\eta+x)}+B_{m,l}e^{il(\eta-x)}\ \ \ (l\neq 0);\\
\psi_{m,0}&=& A_m+B_m\eta.
\eea
Near the singularity, all scalar fields behave like (1+1)-dimensional massless scalar fields on the cylinder labelled by $(\eta,x)$: they oscillate an infinite number of times ($l\in\Zbar_0$) or grow linearly in the conformal time coordinate $\eta$ ($l=0$). Note that despite being singular at the singularity, the wavefunctions are well-defined in each of the four cones, i.e.\ they encode ``matching conditions'' across the singularity.

Now we compute \cite{Berkooz:2002je} the two to two scattering amplitude $1+2\rightarrow3+4$ of closed string tachyons in bosonic string theory (this choice is for technical convenience; similar results are expected for non-tachyonic modes in superstring theory): 
\begin{figure}
\begin{center}
\epsfig{file=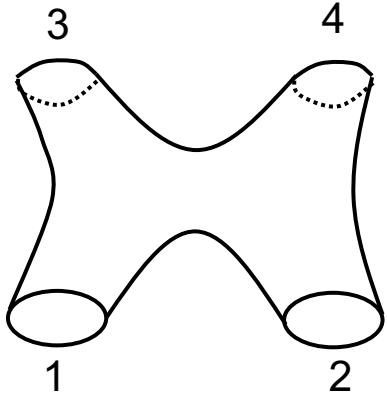, width=7cm}
\end{center}
\caption{Tree-level two to two string scattering amplitude.}\label{fig4pt}
\end{figure}
\be\label{ampw}
\langle \psi_3^*\psi_4^*\psi_1\psi_2\rangle\sim \int dw_1\ldots dw_4 \exp\left(i\sum_{j=1}^4\epsilon_jl_jw_j\right){\cal A},
\ee
where $\epsilon_1=\epsilon_2=1,\, \epsilon_3=\epsilon_4=-1$ and
\be
{\cal A}=\left\langle \prod_{j=1}^4\exp\left[i\left({\epsilon_jm_j\over\sqrt2}X^-e^{-w_j}+{\epsilon_jm_j\over\sqrt2}X^+ e^{w_j}+\epsilon_j\vec p_j\cdot \vec X\right) \right] \right\rangle
\ee
is the standard Virasoro-Shapiro amplitude:
\be
{\cal A}\sim\delta^{(26)}\!\left(\sum_{j=1}^4\epsilon_jp_j\right){\Gamma(-1-{s\over4})\Gamma(-1-{t\over4})\Gamma(-1-{u\over4})\over\Gamma(2+{s\over4})\Gamma(2+{t\over4})\Gamma(2+{u\over4})}.
\ee
Here $s,t,u$ are the Mandelstam variables, given in terms of $v_j\equiv \exp(w_j-w_1)$ as e.g.
\bea
s&=&-(p_1+p_2)^2=-8+m_1m_2(v_2+{1\over v_2})-2\vec p_1\cdot\vec p_2;\\
t&=&-(p_1-p_3)^2=-8-m_1m_3(v_3+{1\over v_3})-2\vec p_1\cdot\vec p_3.
\eea

To evaluate \eq{ampw}, we have to perform four integrals, namely over $w_1$ and $v_j=\exp(w_j-w_1)$. The integral over $w_1$ gives $\delta(\sum \epsilon_il_i)$, expressing momentum conservation along the Milne circle. The integrals over $v_2$ and $v_3$ are algebraic because of the momentum-conserva\-tion delta functions in the $X^\pm$ directions. The remaining integral over $v_4$ has a possible divergence from the large $v_4$ regime, corresponding to
\be
t\approx –(\vec p_1-\vec p_3)^2,\ \ s\approx m_1m_4v_4,
\ee
i.e.\ the Regge limit large $s$, fixed $t$. The large $v_4$ contribution to the amplitude is proportional to
\be
{\Gamma\left[-1+{(\vec p_1-\vec p_3)^2\over4}\right]\over\Gamma\left[2- {(\vec p_1-\vec p_3)^2\over4}\right]}\int^\infty dv_4 v_4^{-{1\over2}(\vec p_1-\vec p_3)^2+i(l_2-l_4)},
\ee
which diverges for sufficiently small momentum transfer $(\vec p_1-\vec p_3)^2\leq2/\alpha^\prime$ (we have restored a factor of $\alpha'$, which we usually set equal to one).

The divergence survives in the ``low energy'' limit $\alpha't\to0$, where it becomes proportional to
\be
{1\over(\vec p_1-\vec p_3)^2}\int^\infty dv_4 v_4^{i(l_2-l_4)}.
\ee
The first factor is a $1/t$ pole from graviton exchange, and in fact the divergence can be precisely reproduced from a dilaton-gravity analysis \cite{Berkooz:2002je}. That analysis shows that the divergence is associated with the region near the singularity. The physical interpretation is that the fast oscillations of wavefunctions near the singularity give rise to a divergent stress tensor, corresponding to an infinite blueshift. This stress energy couples to gravitons and leads to strong gravitational backreaction. One manifestation of this is in the divergence of tree level scattering amplitudes. 

We end this discussion of tree-level divergences and gravitational backreaction with a few comments:
\begin{itemize}
\item
As we have mentioned before, since the first term in an expansion in the string coupling diverges, string perturbation theory breaks down.
\item
Liu, Moore and Seiberg \cite{Liu:2002ft, Liu:2002kb} have analyzed tree-level scattering amplitudes in a supersymmetric time-dependent orbifold with a lightlike singularity \cite{Horowitz:1991ap, Simon:2002ma} and found similar divergences.   
\item
Horowitz and Polchinski \cite{Horowitz:2002mw} have proposed another interpretation of the tree-level divergences. One particle in the Milne orbifold corresponds in the Minkowski covering space to an infinite number of particles (by the method of images). Far images have large relative velocities and cause a large black hole to form, turning the mild-looking orbifold singularity into a more generic spacelike singularity. In terms of Feynman diagrams, the process they consider corresponds to ladder diagrams in which many gravitons are exchanged, i.e.\ high order contributions in the genus expansion.
\item
Our computation shows that  the tree-level divergences are directly related to tree-level exchange of a single graviton. This doesn't exclude a more subtle relation to the non-perturbative process of Horowitz and Polchinski. However, Cornalba and Costa have given an example of a time-dependent orbifold with tree-level divergences and no instability toward formation of a large black hole \cite{Cornalba:2003ze}.
\item
Cornalba and Costa have suggested that the divergences may be cured by working in the eikonal approximation \cite{Cornalba:2003kd}, i.e.\ by resumming an infinite series of ladder diagrams. So far, this suggestion has not been backed up yet by systematic computations, so this is still an interesting open issue.
\end{itemize}

Given that naive string perturbation theory appears to be in poor shape in the presence of cosmological singularities, we have to look for other ways to make progress. 
\begin{figure}
\begin{center}
\epsfig{file=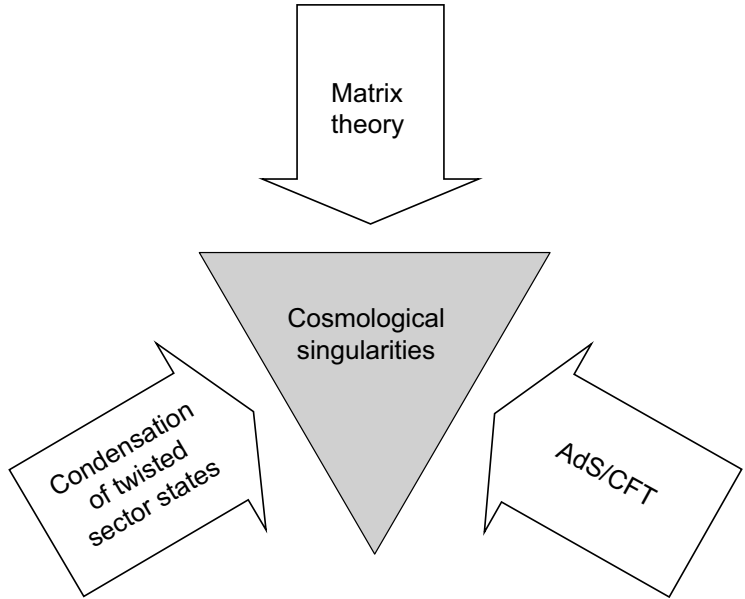, width=12cm}
\end{center}
\caption{How to make progress given that string perturbation theory breaks down?}\label{figProgress}
\end{figure}
One suggestion has been that winding modes, which become light near the singularity, might help resolve the singularity. This will be explored in much of the remainder of the present lecture. Other approaches consist in trying to use non-perturbative formulations of string theory. At the end of this lecture, we will briefly mention a use of the AdS/CFT correspondence. The next two lectures will be devoted to the study of a cosmological singularity in the framework of BFSS matrix theory. 

%%%%%%%%%%%%%%%%%%%%%%%%%%%%%%%
\subsection{Can winding modes resolve the singularity?}

We would like to know whether twisted sector states can resolve the singularity, and in what sense. Let us start with a few remarks:
\begin{itemize}
\item
Tree-level amplitudes diverge, so we cannot be in the situation of static orbifolds, where the singularity was already resolved in string perturbation theory because a condensate of twisted sector modes had ``secretly'' been turned on.
\item
But could it be that time-dependent orbifolds are similar to ALE singularities without B-flux, in that we could identify twisted sector modes whose condensation would lead to a non-singular system?
\item
Even if true, this would at first sight still give a more limited understanding than we have for ALE singularities. For ALE singularities, we know not only a twisted mode (the B-flux) that can be turned on to go to a related non-singular model (the ALE orbifold), but also the light modes (wrapped D-branes) whose inclusion in the effective action resolves the ALE singularity at the "singular" point in moduli space. I would say that the B-flux gives us a way to ``avoid'' the singularity, while the wrapped D-branes allow us to ``understand'' the singularity when it is not avoided. What can we hope for in the case of time-dependent orbifolds?
\item
An important difference with static orbifolds is particle creation. Perhaps particle creation automatically leads to a condensate of twisted sector modes near the singularity?
\end{itemize}

We begin by studying classical configurations of winding strings in the Milne orbifold, following \cite{Berkooz:2004re} (see \cite{Durin:2005ix} for a review). From now on, we allow a parameter $\beta$ in the orbifold identification:
\be
X^\pm\sim e^{\pm2\pi\beta}X^\pm.
\ee
Winding (twisted) strings with winding number $w$ satisfy 
\be
X^\pm(\sigma+2\pi,\tau)=e^{\pm 2\pi w\beta}X^\pm(\sigma,\tau),
\ee
and are explicitly given by
\be
X^\pm(\sigma,\tau)={1\over\nu}e^{\mp\nu\sigma}[\pm\alpha_0^\pm e^{\pm\nu\tau}\mp\tilde\alpha_0^\pm e^{\mp\nu\tau}]+{\rm higher\ oscillators},
\ee
with $\nu=-w\beta$. It turns out that there is a close analogy between winding strings in the Milne orbifold and charged particles in a constant electric field in Minkowski space \cite{Pioline:2003bs}. In particular, twisted string worldsheets can be obtained by smearing trajectories of charged particles under the action of a boost \cite{Berkooz:2004re}. There are two types of such trajectories: those that cross the Rindler horizon and those that stay within a single Rindler wedge. In the Milne orbifold, the former correspond to what are called ``short strings'' and the latter to ``long strings''.

To quantize the winding strings, we impose the commutation relations
\be
[\alpha_0^+,\alpha_0^-]=-i\nu,\ \ [\tilde\alpha_0^+,\tilde\alpha_0^-]=i\nu.
\ee
It turns out \cite{Pioline:2003bs} that these can be identified with the commutation relations for a particle in a constant electric field in Minkowski space, which has the worldline
\be
X^\pm(\tau)=x_0^\pm\pm{a_0^\pm\over\nu}e^{\pm\nu\tau}
\ee
and the commutation relations
\be
[a_0^+,a_0^-]=-i\nu,\ \ [x_0^+,x_0^-]=-{i\over\nu}.
\ee
The identification consists of
\be
a_0^\pm=\alpha_0^\pm,\ \ \ \ x_0^\pm=\mp{\tilde\alpha_0^\pm\over\nu}.
\ee

So let us first discuss the states of a charged particle in a constant electric field, and the phenomenon of pair creation \cite{Pioline:2003bs}. The starting point is the charged Klein-Gordon equation for a particle in a constant electric field. Since the system is invariant under translations in the direction of the electric field, we can diagonalize the corresponding generator. The charged Klein-Gordon equation then takes the form of the time-independent Schr\"odinger equation for an inverted harmonic oscillator in one dimension:
\be
\left(-\partial_u^2-{1\over4}u^2+{M^2\over2\nu}\right)\psi_{\tilde p}(u)=0.
\ee
The worldsheet Hamiltonian has a continuous spectrum of states with arbitrarily negative worldsheet energy. The mass-shell constraint selects one particular worldsheet energy. Tunneling through the potential barrier corresponds to Schwinger pair creation. The Bogoliubov coefficients have been computed; the result is consistent with Schwinger's classic result \cite{Schwinger:1951nm}.

Now we move on to winding modes in the Milne orbifold. The states of a winding string in the Milne orbifold are very similar to those of charged particles in a constant electric field, except that the orbifold projection requires the boost momentum to be integer \cite{Pioline:2003bs}. In the previous analysis, we had diagonalized a generator of spatial translations in Minkowski space. To impose the restriction of integer boost momentum, it is more convenient to diagonalize the boost momentum instead. To accomplish this, one divides Minkowski space in four quadrants and essentially repeats the previous analysis in each quadrant, using Rindler resp.\ Milne coordinates. Pair creation rates of charged particles in the Rindler and Milne regions have been computed, and abundant particle creation has been found \cite{Gabriel:1999yz}. For winding strings in the Milne orbifold, the interpretation of these creation rates is less clear. In particular, it is unclear how one should define a vacuum state for the long strings, which have infinite mass and live in a region with closed timelike curves \cite{Berkooz:2004re}. 

From the point of view of singularity resolution it is important to ask how the produced winding strings backreact on the geometry. Taking into account the effect of correlated pairs of produced winding strings seems beyond what can be done using current technology. A related problem that seems more tractable is to study the effect of a classical condensate of winding strings. The hope would be that such a condensate might lead to a bounce instead of the Milne singularity, which would give hope that the pair produced winding strings could smooth out the singularity as well \cite{Berkooz:2004re}. To study this backreaction quantitatively, one needs to compute correlation functions involving twisted fields. Imagine adding a condensate of marginal twist operators to the free worldsheet action of the Milne orbifold:
\be
S_\lambda=\int d^2\sigma \partial_\alpha X^+ \partial^\alpha X^-
+\lambda_{-w}V_w+\lambda_wV_{-w}.
\ee
This gives rise to a one-point function for untwisted fields,
\be
\langle V_U \rangle_\lambda\sim\lambda_w\lambda_{-w}\langle w|\,V_U\,|-w\rangle,
\ee
which should be compensated by deforming the untwisted action. This three point function has been computed \cite{Berkooz:2004yy}. Some correlators with more than two twist fields have also been computed by these authors, using recent results on the Wess-Zumino-Witten model of a 4d Neveu-Schwarz plane wave \cite{D'Appollonio:2003dr, Cheung:2003ym}. This work has not led to a definite conclusion yet. 

Recently, a variation of the Milne orbifold with localized closed string tachyons has received a lot of attention. So far in this lecture, we have done our computations for the bosonic string, with the understanding that they would be similar for the superstring, which is tachyon-free. An interesting new ingredient is introduced \cite{McGreevy:2005ci} when one considers a variation of the Milne orbifold where spacetime fermions are antiperiodic around the Milne circle and where the ``wrong'' GSO-projection is used for states with odd winding number: twisted sector tachyons appear when the length of the Milne circle is smaller than the string length. If the parameter $\beta$ of the boost identification $X^\pm\sim\exp(\pm2\pi\beta)X^\pm$ is sufficiently small, the tachyon wave function grows exponentially as one goes back in time towards the big bang. In the Milne regions, the tachyon is localized near the singularity. 

What is the effect of an exponentially large tachyon condensate $T$ near the Milne singularity? The tachyon appears in the string worldsheet as 
\be
S=\int d\tau d\sigma \left(-G_{\mu\nu}(X)\partial_\alpha X^\mu\partial^\alpha X^\nu-T(X)\right).
\ee
Comparing with the worldline action of a relativistic particle
\be
S=\int d\tau \left(-G_{\mu\nu}(X)\dot X^\mu\dot X^\nu- m^2\right),
\ee
we see that a large tachyon condensate makes all closed string states very heavy. McGreevy and Silverstein \cite{McGreevy:2005ci} conclude that because of the tachyon condensate, all closed string degrees of freedom, including gravity itself, will be lifted before the singularity is reached. Furthermore, they claim (based on analytic continuations of results in Liouville theory) that perturbative string amplitudes are dominated by the region far away from the singularity. The tachyon condensate is interpreted as a ``nothing'' phase from which spacetime emerges. Perturbative string theory is claimed to be valid everywhere.

This picture raises a number of interesting questions. First, analytic continuations tend to be very subtle in time-dependent backgrounds, especially beyond tree level -- is it possible to confirm this picture with more explicit computations? Second, is there a non-perturbative interpretation of the nothing phase? Third, the tachyon condensate is similar to the B-field in static orbifolds, in that turning on this twisted mode appears to avoid the singularity. One could still ask about the physics of the singularity without the tachyon turned on (the analogue of understanding ALE singularities without B-flux): how is the singularity resolved when it is not avoided?

Finally, we mention some interesting attempts to determine the backreaction using D-brane probes. To study a spacetime at substringy length scales, one probes the spacetime using D-branes \cite{Douglas:1996yp}. The spacetime appears as (a branch of) the moduli space of the D-brane effective field theory. To use this technique for spacetimes with cosmological singularities, one uses D-instanton probes. This has been applied \cite{Berkooz:2005ym} to the parabolic orbifold, the time-dependent orbifold with a lightlike singularity studied in \cite{Horowitz:1991ap, Simon:2002ma, Liu:2002ft, Liu:2002kb}. The interesting question is how the moduli space (and thus the spacetime) is deformed when a twisted sector mode condenses. Some evidence has been presented that the parabolic orbifold flows to a well-understood orbifold \cite{Berkooz:2005ym}. This technique has also been used for the variation of the Milne orbifold with localized closed string tachyons. However, the analysis is subtle and two groups arrive at contradictory conclusions as to whether the big bang region is connected to a big crunch region after tachyon condensation \cite{She:2005mt, Hikida:2005xa}.

%%%%%%%%%%%%%%%%%%%%%%%%%%%%%%%
\subsection{Using the AdS/CFT correspondence}

The AdS/CFT correspondence (see \cite{Aharony:1999ti} for a review) states that type IIB string theory on $AdS_5\times S^5$ with string coupling $g_s$ and radius of curvature $R$ is dual (equivalent) to ${\cal N}=4\ SU(N)$ super-Yang-Mills theory with Yang-Mills coupling $g_{YM}$ on the (3+1)-dimensional boundary of $AdS_5$. The AdS/CFT ``dictionary'' identifies $g_s$ with $1/N$, and $RM_s^4$ with the 't Hooft coupling $g_{YM}^2N$, with $M_s$ the string mass (i.e.\ the square root of the string tension).

The AdS/CFT correspondence gives a non-perturbative definition of string theory in (asymptotically) anti-de Sitter space. The weak coupling limit of the string theory corresponds to the large $N$ limit of the gauge theory. The weak curvature (supergravity) limit of the string theory corresponds to the strong 't Hooft coupling limit of the gauge theory.

The correspondence can be used to study a spacelike singularity by considering an AdS Schwarzschild black hole. 
\begin{figure}
\begin{center}
\epsfig{file=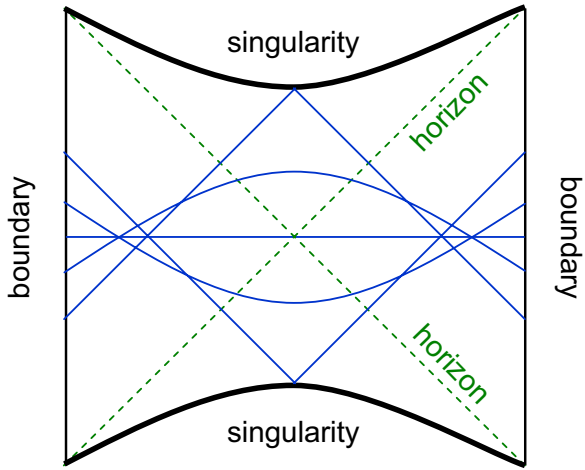, width=10cm}
\end{center}
\caption{Penrose diagram of an AdS Schwarzschild black hole, and some geodesics.}\label{figBH}
\end{figure}
A Schwarzschild black hole in five-dimensional anti-de Sitter space is mapped via the AdS/CFT correspondence to an entangled state in the tensor product of two $N=4$ super-Yang-Mills theories (one at each asymptotically AdS boundary of the black hole spacetime) \cite{Maldacena:2001kr}. Analytically continued versions of the correlation functions get contributions from spacelike  geodesics that come close to the singularity; their divergences have been interpreted as a subtle signature of the black hole singularity. The goal is to use gauge theory computations to determine if and how the singularity is resolved in the full, non-perturbative string theory. This is challenging because the gauge theory is strongly coupled in the regime of interest, and because the signature of the singularity is very subtle. For details, we refer to \cite{Kraus:2002iv, Fidkowski:2003nf, Festuccia:2005pi}. See \cite{Hertog:2005hu} for a different use of the AdS/CFT correspondence in the study of cosmological singularities.

To summarize this second lecture, we have seen that two to two scattering amplitudes of untwisted perturbations in the Milne orbifold exhibit unusual divergences. These divergences are associated with large tree-level gravitational backreaction and lead to the breakdown of the string perturbation expansion. One hope is that cosmologically produced winding strings might backreact on the geometry in such a way as to avoid the Milne singularity. Despite concrete progress in our understanding of winding strings in the Milne orbifold, we don't know if this hope is realized. In a variation of the model with localized closed string tachyons, it has been claimed that a tachyon condensate phase might replace the cosmological singularity. This is an interesting idea, but more explicit computations seem to be required to settle whether such a ``nothing'' phase is indeed realized. If so, it would still be interested to understand the physics of the singularity in the case that no tachyon has been condensed. One non-perturbative approach to the study of spacelike singularities uses the AdS/CFT correspondence to study the singularity of an AdS-Schwarzschild black hole. While a subtle signature of the singularity has been found, it is hard to obtain detailed information about the local physics of the singularity from the point of view of an observer at infinity.

%%%%%%%%%%%%%%%%%%%%%%%%%%%%%%%%%%%%%%%%%%%%%%%%%%%%%%%%%%%%%%%%%%%%%%%%%%%%%%%%%%%%%%%%%%%%%%%%%%%%%%%%%%%%%%%%%%%%%%%%%%%%%%%%%%%%%%%%%%%%%%%%%%%%%%%%%%%%%%%%%%%%%%%%%%%%%%%%%%%%%%%%%%%%%%%%%%%%%%%
\section{Lecture III: Light-like linear dilaton and matrix theory}

In this and the following lecture, we will discuss a model with a cosmological singularity \cite{Craps:2005wd} in the framework of BFSS matrix theory \cite{Banks:1996vh}, which was originally developed as a non-perturbative description of string theory in (asymptotically) Minkowski space. Earlier cosmological applications of matrix theory are \cite{Alvarez:1997fy, Freedman:2004xg}. Cosmology in the context of the ``old'' matrix model has also been actively studied, starting with \cite{Karczmarek:2003pv}.

%%%%%%%%%%%%%%%%%%%%%%%%%%%%%%%%
\subsection{Light-like linear dilaton}

The starting point is an extremely simple time-dependent solution of ten-dimensional (type IIA) string theory, namely flat space with a lightlike linear dilaton (it preserves half of the supersymmetries):
\bea
ds_{10}^2&=&-2dX^+dX^-+(dX^i)^2\nonumber\\
\Phi&=&-QX^+.\label{LLD}
\eea
The dilaton $\Phi$ is a scalar field that appears in the low-energy effective action as
\be
S\sim\int d^{10}x\,\sqrt G\, e^{-2\Phi}\left(R+4\partial_\mu\Phi\partial^\mu\Phi+\ldots\right).
\ee
Therefore, the exponential of the dilaton can be viewed as the string coupling ``constant'':
\be
g_s=e^\Phi.
\ee
\begin{figure}
\begin{center}
\epsfig{file=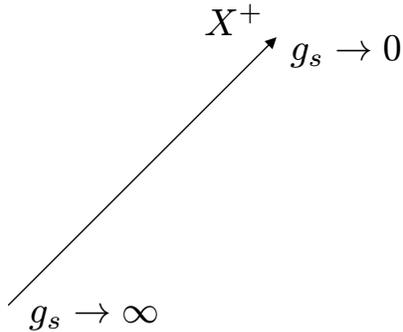, width=7cm}
\end{center}
\caption{The lightlike linear dilaton background: the string coupling evolves from weak to strong as a function of lightcone time.}\label{figLLD}
\end{figure}

In the Einstein conformal frame, where the dilaton factor in front of the Ricci scalar is absent, the metric is non-trivial and exhibits a big bang singularity for $X^+\to-\infty$:                   \be
ds^2_E=e^{QX^+/2}\left[-2dX^+dX^-+(dX^i)^2\right].
\ee    
Introducing a new coordinate $u=\exp(QX^+/2)$, the metric can be rewritten as
\be
ds_E^2=-{4\over Q}du\,dX^-+u(dX^i)^2,
\ee
from which we find the following non-vanishing components of the Riemann curvature tensor:
\be
R_{iuiu}={1\over4u}.
\ee
\begin{figure}
\begin{center}
\epsfig{file=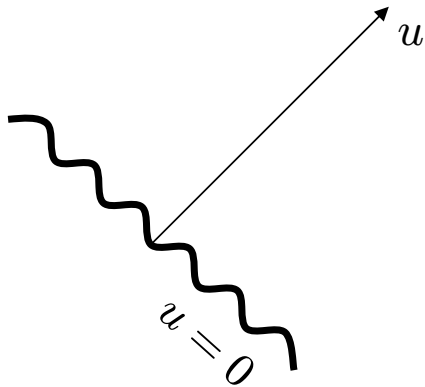, width=7cm}
\end{center}
\caption{The lightlike linear dilaton background in Einstein frame.}\label{figLLDu}
\end{figure}

The main difference between the singularity at $u=0$ (corresponding to infinite string coupling) and a ``standard'' big bang singularity is that the former is lightlike, not spacelike. However, in the context of perturbative string theory we have seen that lightlike singularities were about just as hard to treat as spacelike ones. Therefore, a string theory resolution of any cosmological singularity, lightlike or spacelike, would be great progress.

Later in these lectures, we will see in detail how the above spacetime can be studied in matrix theory. Now I will first give a brief introduction to matrix theory in flat spacetime. 
%%%%%%%%%%%%%%%%%%%%%%%%%%%%%%
\subsection{Matrix theory in flat spacetime}

Let us first recall the correspondence between type IIA string theory and M-theory. When studying type IIA string theory perturbatively, we compute asymptotic series in $g_s$, which are useful for very small values of $g_s$. The question is what happens when $g_s$ is large rather than small. An important tool in addressing this question is supersymmetry, in particular the fact that masses of BPS states are protected against quantum corrections. For instance, for any value of $g_s$, the mass of a D0-brane is 
\be
\tau_{D0}={1\over g_s\sqrt{\alpha^\prime}},
\ee
which implies that D0-branes become light at strong coupling. It is also known that there exist bound states of $N$ D0-branes with masses $N\tau_{D0}$. This matches the spectrum of Kaluza-Klein (KK) modes for a periodic dimension of radius $R_{11}=g_s\sqrt{\alpha^\prime}$. It has been conjectured \cite{Witten:1995ex} that ten-dimensional type IIA string theory is a circle compactification of an eleven-dimensional theory, called ``M-theory''. The low energy effective field theory for M-theory is eleven-dimensional supergravity. The relation between the eleven-dimensional metric and ten-dimensional fields is as follows:
\bea
ds_{11}^2&=&G^{11}_{MN}(x^\mu)dx^Mdx^N\\
&=&\exp\left(-{2\over3}\phi(x^\mu)\right)G_{\mu\nu}^{10}dx^\mu dx^\nu
+\exp\left({4\over3}\phi(x^\mu)\right)\left[dx^{10}+C_\nu(x^\mu)dx^\nu\right]^2,\nonumber
\eea
where $\phi$ is the dilaton, $G_{\mu\nu}^{10}$ is the ten-dimensional metric and $C_\nu$ is the Ramond-Ramond one-form potential. Dimensional reduction keeps only modes with $p_{10}=0$. The KK modes with nonzero $p_{10}$ correspond to D0-branes and their bound states. 

The question now is: what is M-theory? We know what it is when compactified on a small circle and we know its low-energy limit, but we would like to have a microscopic description of the theory. Matrix theory is a non-perturbative description of M-theory in eleven-dimensional Minkowski space and some of its compactifications.

A way to derive matrix theory \cite{Seiberg:1997ad} is by performing a Discrete Light-Cone Quantization (DLCQ) of M-theory \cite{Susskind:1997cw}. Consider the time coordinate $t$ and one of the space coordinates $x$, and make a periodic lightlike identification,
\be\label{identification}
\left( \begin{array}{c} x  \\ t\end{array} \right) \sim
\left( \begin{array}{c} x-{R/\sqrt2} \\ t+{R/\sqrt2}\end{array} \right),
\ee
i.e.\ $x^-\sim x^-+R$. The momentum conjugate to $x^-$ is now quantized: $p^+=n/ R$ with integer $n$ (we are suppressing a factor of $2\pi$). In fact, in (discrete) lightcone quantization all quanta have nonnegative $n$, and the modes with $n=0$ are non-dynamical and can be integrated out. We can thus restrict to $n>0$. See for instance \cite{Bigatti:1997jy} for more details. In discrete lightcone quantization, one focuses on a sector with fixed total lightcone momentum 
\be
p^+={N\over R}
\ee
with $N>0$. In other words, one studies the theory sector by sector, where each sector is labeled by the conserved quantum number $N$. 

How does one define a theory with a lightlike direction compactified? This can be done by considering the lightlike compactification as a limit of a spacelike compactification \cite{Seiberg:1997ad}:
\be\label{idlimit}
\left( \begin{array}{c} x  \\ t\end{array} \right) \sim \left( \begin{array}{c} x-\sqrt{{R^2\over2}+R_s^2} \\ t+{R/\sqrt2}\end{array} \right)
\ee
with $R_s\to 0$, $R_s$ being the proper size of the spacelike circle.

The Lorentz boost
\be
\left( \begin{array}{c} x ^\prime \\ t^\prime\end{array} \right) =\left( \begin{array}{cc}\cosh\beta&\sinh\beta\\ \sinh\beta&\cosh\beta\end{array}\right) \left( \begin{array}{c} x\\ t\end{array} \right)
\ee
with 
\be
\cosh\beta=\sqrt{1+{R^2\over2R_s^2}}
\ee
of course preserves the form of the Minkowski metric while turning the identification into
\be
\left( \begin{array}{c} x^\prime  \\ t^\prime\end{array} \right) \sim \left( \begin{array}{c} x^\prime-R_s \\ t^\prime\end{array} \right).
\ee
We thus see that M-theory on a lightlike circle is equivalent to M-theory on a spacelike circle with radius $R_s\to 0$.

But that is the same as type IIA string theory with \cite{Witten:1995ex}
\be
g_s=(R_sM_p)^{3/2}, \ \ \ \ {1\over\sqrt{\alpha^\prime}}\equiv M_s=\sqrt{R_sM_p^3}, 
\ee
with $M_p$ the eleven-dimensional Planck mass. In the $R_s\to 0$ limit, we get weakly coupled type IIA strings ($g_s\to 0$), but the string length becomes large ($\alpha'\to\infty$), which would seem problematic. However, it doesn't make sense to say that a dimensionful quantity such as $\alpha'$ is small or large: one should always talk about dimensionless ratios being small or large. In M-theory on a lightlike circle, we are interested in states with lightcone momentum $p^+=N/R$ and lightcone energy $P^-$, determined by the physical process we wish to study. After the boost, we are considering states with momentum $P'=N/R_s$ (i.e.\ we are in a sector with $N$ D0-branes) and energy $E'=N/R_s+\Delta E'$. Since
\be
P^-={1\over\sqrt2}(E-P)={1\over\sqrt2}e^\beta(E^\prime-P^\prime)\approx{R\over R_s}\Delta E^\prime,
\ee
we find
\be
\Delta E^\prime\approx{R_sP^-\over R}\    \Rightarrow\ {\Delta E^\prime\over M_s}\approx {\sqrt{R_s}P^-\over RM_p^{3/2}}\to0.
\ee
So the energies of interest are actually small compared to the string mass, or in other words, the string length goes to zero compared to the length scales of interest.

So M-theory on a lightlike circle with radius $R$ in a sector with $p^+=N/R$ is equivalent to type IIA string theory in the presence of $N$ D0-branes with 
\be\label{YMlimit}
g_s\to0,\ \sqrt{\alpha^\prime}\Delta E^\prime\to0.
\ee
In this limit, where we keep the momenta in the non-compact directions fixed, the only non-decoupled degrees of freedom are the $N\times N$ matrices of the D0-brane worldvolume theory, which in the low-energy limit \eq{YMlimit} is the dimensional reduction of 10d super-Yang-Mills theory, known as ``matrix theory'' \cite{Banks:1996vh}. See \cite{Polchinski:1999br} for a nice discussion of this limit from various points of view.

Eventually, one wants to decompactify the lightlike circle:
\be
R\to\infty,\ N\to\infty\ {\rm with}\ p^+=N/R\ {\rm fixed}.
\ee
This removes the lightlike ``box'', while keeping the physical lightcone momentum fixed. 

A complication in DLCQ of field theories is that the zero modes (modes with $p^+=0$) are strongly coupled, since $R_s$ appears in the zero mode action as a multiplicative factor \cite{Hellerman:1997yu}. M-theory appears to be better behaved in this respect: we know that M-theory on a small circle is a weakly coupled (string) theory. See \cite{Polchinski:1999br, Banks:1999az} for further discussion.

To summarize, the DLCQ of M-theory in a sector with $N$ units of lightcone momentum is given by the low-energy limit of the worldvolume theory of $N$ D0-branes. This is the dimensional reduction of (9+1)-dimensional super-Yang-Mills theory to 0+1 dimensions: matrix quantum mechanics. To get uncompactified M-theory, one has to take a large $N$ limit.  
 
In the previous discussion, we started from eleven-dimensional M-theory, compactified the $x^-$ direction and obtained the worldvolume theory of $N$ D0-branes in type IIA string theory, which is M-theory with M-theory circle along $x^-$ with radius $R_s$. Now we compactify both sides of this correspondence on a circle along the $x^9$ direction with radius $R^9$. 
\begin{figure}
\begin{center}
\epsfig{file=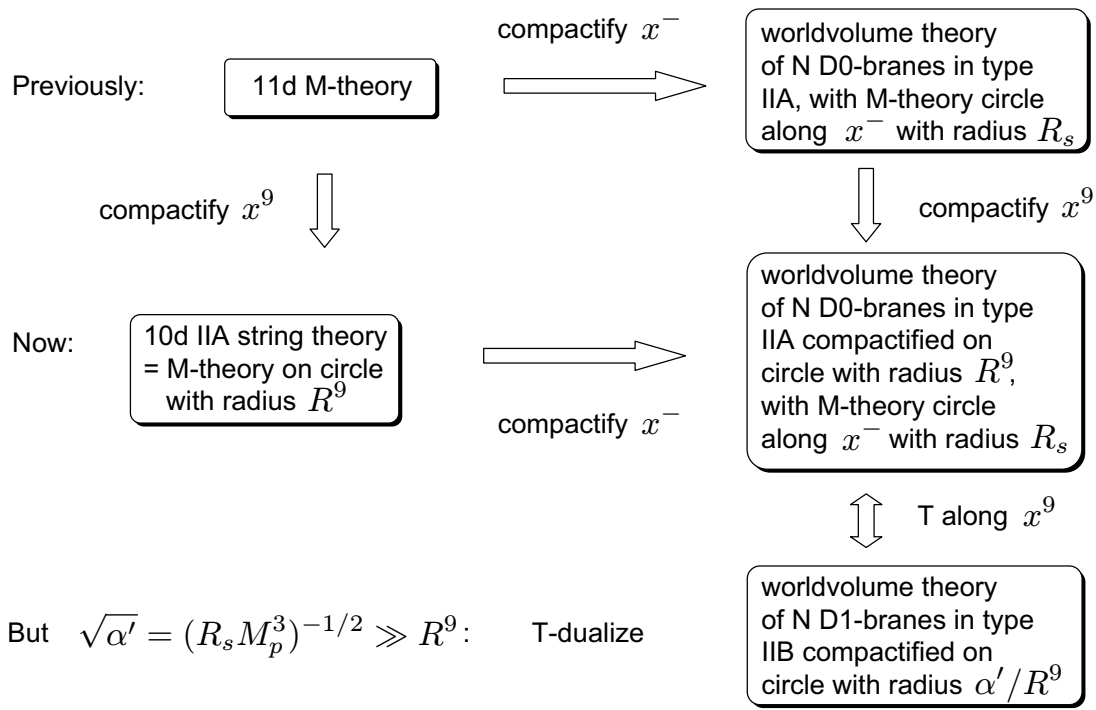, width=16.3cm}
\end{center}
\caption{Matrix description of type IIA strings: matrix string theory.}\label{figMatrixString}
\end{figure}
The starting point then becomes M-theory on a circle with radius $R^9$, which is the same as type IIA string theory. The result becomes the worldvolume theory of $N$ D0-branes in type IIA string theory compactified on a circle with radius $R^9$, where the type IIA string theory is M-theory with M-theory circle along $x^-$ with radius $R_s$. But in the limit $R_s\to 0$, the string length $\sqrt{\alpha^\prime}=(R_sM_p^3)^{-1/2}$ is much larger than $R^9$ so it is convenient to perform a T-duality along the $x^9$ direction. Thus we obtain the worldvolume theory of $N$ D1-branes in type IIB string theory compactified on a circle with radius $\alpha'/R^9$. This is known as ``matrix string theory'' \cite{Motl:1997th, Banks:1996my, Dijkgraaf:1997vv}.

In the previous derivation, the original type IIA string theory (with $N$ units of lightcone momentum) was related to an auxiliary type IIA string theory (with $N$ units of D0-brane charge) by a 9-11 flip (i.e.\ by viewing two different circles as the M-theory circle). The 9-11 flip is equivalent to a sequence of T-duality, S-duality and T-duality:
\be
{\rm momentum}  \stackrel{T}{\rightarrow}  {\rm F1\ winding}   \stackrel{S}{\rightarrow}     {\rm D1\ winding}   \stackrel{T}{\rightarrow}  {\rm D0\ charge}
\ee
Thus the original type IIA theory is related to the auxiliary type IIB theory above by a sequence of T-duality and S-duality.

So matrix string theory is a non-perturbative formulation of type IIA superstring theory in (9+1)-dimensional Minkowski space. It is described by the low-energy effective action of $N$ D1-branes in type IIB string theory, which is ${\cal N}=8$ super-Yang-Mills theory in 1+1 dimensions with gauge group $U(N)$, in a large $N$ limit:
\be\label{matrixstringaction}
S=\int d\tau d\sigma\, {\rm Tr}\left((D_\mu X^i)^2+\theta^TD\!\!\!\!\slash\ \theta+g_s^2F_{\mu\nu}^2-{1\over g_s^2}[X^i,X^j]^2 +{1\over g_s}\theta^T\gamma_i[X^i,\theta]
\right).
\ee
The fields $X^i, \theta^\alpha, \theta^{\dot\alpha}$ are $N\times N$ Hermitean matrices transforming in the ${\bf 8}_v, {\bf 8}_s, {\bf 8}_c$ representations of the $SO(8)$ R-symmetry group of transverse rotations. The worldsheet is an infinite cylinder with coordinates $(\tau,\sigma)$, where $\sigma\sim\sigma+2\pi$. (We are setting $\ell_s\equiv\sqrt{\alpha'}=1$ except when we explicitly wish to display the $\ell_s$-dependence.) These are the same fields as in the lightcone Green-Schwarz formulation of the superstring, except that now they are matrix-valued. The relation is that eigenvalues of the matrices $X^i$ correspond to coordinates of (pieces of) superstring, and similarly for the fermions. Matrix string theory is a second quantized, non-perturbative extension of the Green-Schwarz superstring.

To summarize, the DLCQ of type IIA string theory in a sector with $N$ units of lightcone momentum is given by the low-energy limit of the worldvolume theory of $N$ D1-branes in type IIB string theory. This is the dimensional reduction of (9+1)-dimensional super-Yang-Mills theory to 1+1 dimensions. To get uncompactified type IIA string theory, one has to take a large $N$ limit.  
%%%%%%%%%%%%%%%%%%%%%%%%%%%%%%
\subsection{A first look at matrix theory in the lightlike linear dilaton background}

In the next lecture, we will generalize the derivation of matrix string theory to the lightlike linear dilaton background. The result will be as follows. The time coordinate $\tau$ appearing in the action \eq{matrixstringaction} is related to the spacetime lightcone time coordinate $X^+$ by $X^+=\tau/R$, where $R$ is the parameter of the lightlike identification \eq{identification}. (In these lectures, we will often omit factors of 2 and $\pi$ when they are not essential; the reader interested in those factors is referred to the original papers \cite{Craps:2005wd, Craps:2006xq}.) The result of the analysis of the next lecture is that we can simply plug $g_s=\exp(-QX^+)=\exp(-Q\tau/R)$ into the action \eq{matrixstringaction}. This gives rise to (1+1)-dimensional super-Yang-Mills theory on the cylinder, with Yang-Mills coupling
\be\label{coupling}
g_{YM}={1\over\ell_s}\exp\left({Q\ell_s\tau\over R}\right).
\ee
This theory is equivalent to (1+1)-dimensional super-Yang-Mills theory with constant coupling on the future cone of the Milne orbifold,
\be
ds^2=e^{2Q\tau\over R}(-d\tau^2+d\sigma^2),
\ee
which we discussed in the previous lectures in a different context (as part of a string theory spacetime).

In the first description, in terms of super-Yang-Mills theory on a cylinder, cosmological evolution from early to late times corresponds to evolution from weak to strong Yang-Mills coupling. In the second description, in terms of super-Yang-Mills theory on the Milne orbifold, cosmological evolution corresponds to evolution from a small to a big space, i.e.\ from the ultraviolet limit of the field theory to the infrared limit. In this sense, cosmological evolution can be thought of as ``renormalization group flow''. 
\begin{figure}
\begin{center}
\epsfig{file=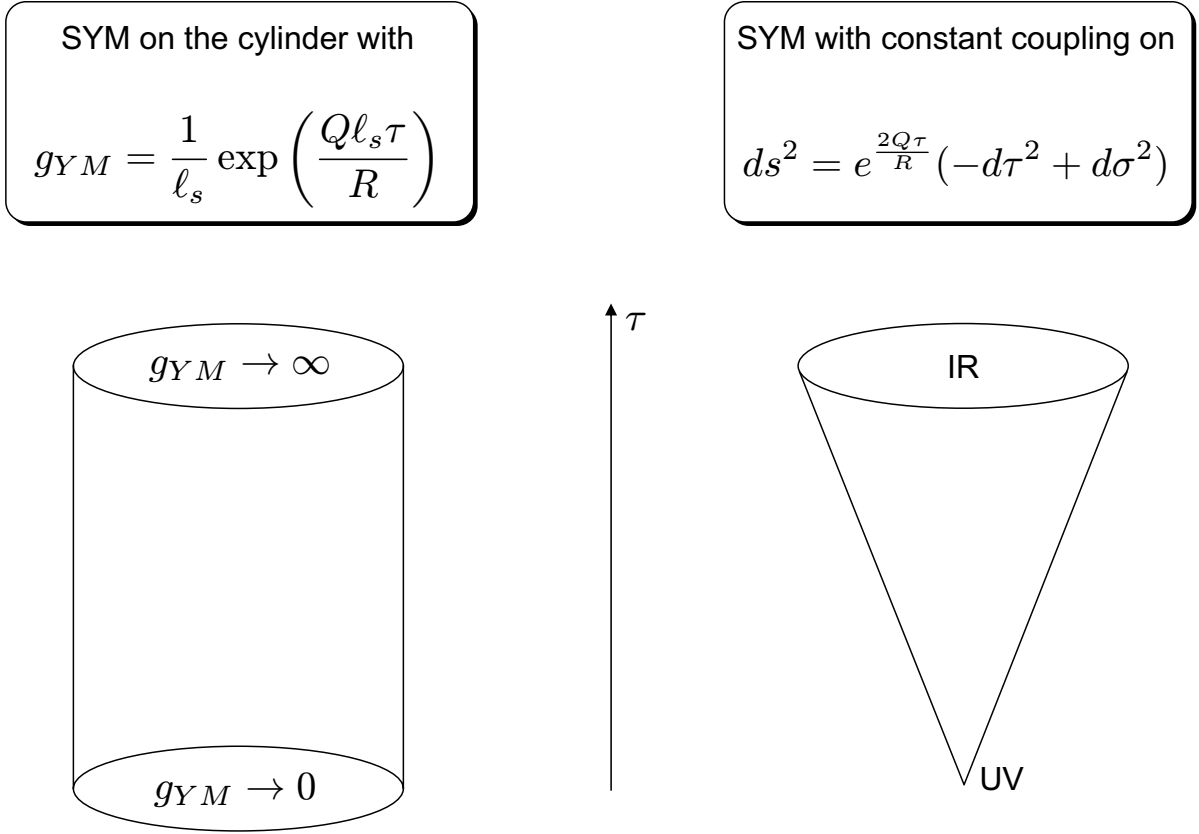, width=16.3cm}
\end{center}
\caption{Cosmological evolution as ``renormalization group flow''.}\label{figRG}
\end{figure}

What is the relation between this (1+1)-dimensional Yang-Mills theory and the ten-dimensional Minkowski spacetime we started with? 
\begin{figure}
\begin{center}
\epsfig{file=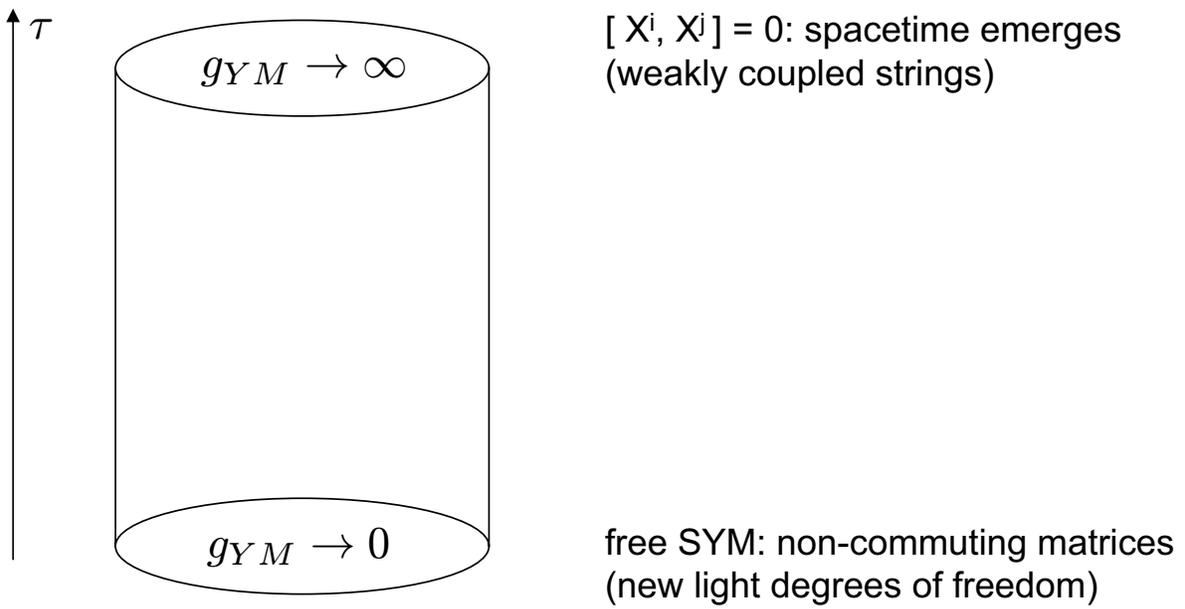, width=16.3cm}
\end{center}
\caption{Emergence of spacetime in the matrix big bang model.}\label{figEmergence}
\end{figure}
At late times, the Yang-Mills coupling \eq{coupling} is strong and the coefficient of the commutator squared potential in \eq{matrixstringaction} is large. This forces the matrices $X^i$ to commute and leads to a spacetime picture just as in flat space matrix theory: the eigenvalues of the commuting matrices correspond to spacetime positions, whereas the off-diagonal matrix elements are heavy and can be integrated out. However, at early times the potential turns off and we need to consider the full, non-diagonal matrices. The off-diagonal matrix elements provide additional light degrees of freedom, and there is no good spacetime description.
Thus the region near the big bang singularity (in Einstein frame) of the original spacetime is replaced by a theory of non-commuting matrices. We call this model the ``matrix big bang''. 

%%%%%%%%%%%%%%%%%%%%%%%%%%%%%%%%%%%%%%%%%%%%%%%%%%%%%%%%%%%%%%%%%%%%%%%%%%%%%%%%%%%%%%%%%%%%%%%%%%%%%%%%%%%%%%%%%%%%%%%%%%%%%%%%%%%%%%%%%%%%%%%%%%%%%%%%%%%%%%%%%%%%%%%%%%%%%%%%%%%%%%%%%%%%%%%%%%%%%
\section{Lecture IV: Matrix big bang}

In this lecture, we first derive the matrix big bang model using discrete lightcone quantization. Then we discuss some quantum effects in this model.
%%%%%%%%%%%%%%%%%%%%%%%%%%%%%%
\subsection{Derivation of the matrix big bang via DLCQ}

We start with the type IIA string background \eq{LLD},
\bea
ds_{10}^2&=&-2dX^+dX^-+(dX^i)^2\nonumber\\
\Phi&=&-QX^+,
\eea
and perform discrete lightcone quantization, identifying $X^-\sim X^-+R$ and focusing on a sector with lightcone momentum $p^+=N/R$. Eventually, we want to take the limit $N\to\infty,\, R\to\infty$ with $p^+$ fixed. 

As before, we define this lightlike compactification as a limit of a spacelike compactification \cite{Craps:2005wd}. However, because the dilaton depends on $X^+$, we cannot impose the identification \eq{idlimit}, since it is not a symmetry of the background. Instead, we single out one of the space directions, say $X^1$, and impose the identification
\be
\left( \begin{array}{c} X^+\\ X^-\\X^1  \end{array} \right)\sim \left( \begin{array}{c} X^+\\ X^-\\X^1  \end{array} \right)+\left( \begin{array}{c} 0\\ R\\ \epsilon R  \end{array} \right),
\ee 
where the lightlike limit corresponds to $\epsilon\to0$.
 
The Lorentz transformation
\bea\label{Lorentz}
X^+&=&\epsilon x^+\\
X^-&=&{x^+\over 2\epsilon}+{x^-\over\epsilon}+{x^1\over\epsilon}\nonumber\\
X^1&=&x^++x^1\nonumber
\eea 
leads to the background
\bea
ds_{10}^2&=&-2dx^+dx^-+(dx^i)^2\\
\Phi&=&-\epsilon Qx^+
\eea 
with identification $x^1\sim x^1+\epsilon R$ and momentum quantization condition $p^1=N/\epsilon R$. 
 
T-duality along $x^1$ and S-duality leads to type IIB string theory with $N$ D1-branes wrapped around $x^1$ in the background
\bea\label{IIBbackground}
ds^2&=&\epsilon R\, e^{\epsilon Qx^+}\left[-2dx^+dx^-+(dx^i)^2\right]\\
\Phi&=&\epsilon Qx^++\log\epsilon R\nonumber
\eea
with identification $x^1\sim x^1+1/ \epsilon R$.

The dynamics of $N$ D1-branes in the background \eq{IIBbackground} is described by (a non-abelian version of) the Dirac-Born-Infeld action
\be\label{DBI}
S_{D1}=-\int d\tau d\sigma\,e^{-\Phi}\sqrt{-\det(\partial_\alpha X^\mu
\partial_\beta X^\nu G_{\mu\nu}+F_{\alpha\beta})}.
\ee
A ground state configuration of the D1-branes is obtained by aligning them with the $(t,x^1)$ plane in the spacetime \eq{IIBbackground}, where $t=(x^++x^-)/\sqrt2$. 
\begin{figure}
\begin{center}
\epsfig{file=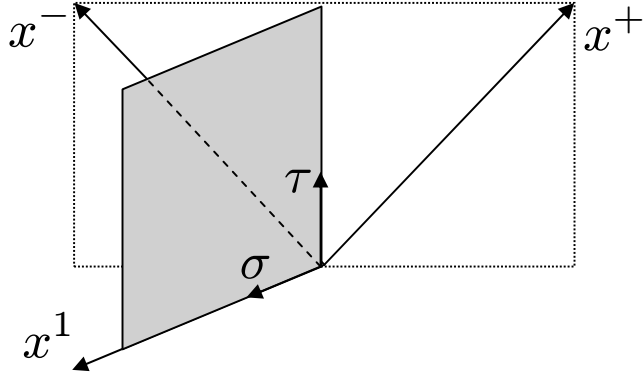, width=10cm}
\end{center}
\caption{A D1-brane ground state in the background \eq{IIBbackground}.}\label{figGroundState}
\end{figure}
We make the gauge choice (choice of worldvolume coordinates $\sigma, \tau$)
\be\label{xtau}
x^1={\sigma\over\epsilon R}\ ,\ \ x^+={\tau\over\epsilon R}
\ee 
with $\sigma\sim\sigma+2\pi$. Introducing new fields $y^j$,
\be
x^-={\tau\over\epsilon R}+y^1,\ \ x^i=y^i\ \ \ (i=2,\ldots 8),
\ee 
the action \eq{DBI} becomes
\bea\label{DBIbis}
S_{D1}&=&\int d\tau d\sigma\, {\rm Tr}\left\{(\partial_\alpha y^i)^2
+{1\over g_{YM}^2}F_{\alpha\beta}^2-g_{YM}^2[y^i,y^j]^2
\right\}\nonumber\\
&&+\ {\rm fermions}\ +\ {\rm tension}\ +\ {\rm higher\ derivatives}
\eea
with
\be
g_{YM}={1\over \ell_s}\exp\left({Q\tau\ell_s\over R}\right)={1\over \ell_s g_s^{IIA}}.
\ee
Note that the symmetry between the first and the other seven transverse directions has been restored in the terms displayed. The tension in \eq{DBIbis} is field-independent and can be ignored. If we can also ignore the higher derivative terms, i.e.\ if we are allowed to restrict to the low-energy limit of the Dirac-Born-Infeld action, then \eq{DBIbis} reduces precisely to the matrix string action \eq{matrixstringaction} with the time-dependent string coupling plugged in as at the end of the previous lecture.

We now argue that it is consistent to truncate type IIB string theory with $N$ D1-branes wrapped around $x^1$ in the background \eq{IIBbackground} to the low-energy limit of the Dirac-Born-Infeld action \eq{DBIbis}. What we have to show is that closed strings and massive open strings decouple from the low-energy dynamics on the branes. If in the original type IIA background \eq{LLD} we focus on a perturbation with lightcone energy $E^-$, defined with respect to the lightcone time coordinate $X^+$, then in the dual type IIB background \eq{IIBbackground} we are focusing on lightcone energies $\epsilon E^-$ with respect to $x^+$.% 
\footnote{
To make this more precise, first note that the gauge choice \eq{xtau} implies that fluctuations in the momenta conjugate to $x^+$ and $x^1$ vanish: $\delta\tilde p^+=\delta\tilde p^1=0$, where a tilde denotes a momentum with respect to the $(x^+, x^-, x^i)$ coordinates. If we consider a fluctuation $\delta\tilde p^-\equiv\tilde E^-$ in the lightcone energy, the Lorentz transformation \eq{Lorentz} tells us that the fluctuations in the momenta conjugate to the $(X^+, X^-, X^i)$ coordinates are $\delta p^-\equiv E^-=\tilde E^-/\epsilon$ and $\delta p^+=\delta p^1=0$. This argument gives the announced relation $\tilde E^-=\epsilon E^-$, and moreover shows that not only fluctuations in the lightcone energy but also fluctuations in the usual energy in the $(x^+, x^-, x^i)$ frame are of order $\epsilon$. 
}
Since only dimensionless ratios of lengths are physical, the spacetime \eq{IIBbackground} with constant string length $\ell_s$ is equivalent to Minkowski space with time-dependent string length
\be
\ell_s^{\rm eff}={\ell_s^{3/2} \over\sqrt{\epsilon R}} \exp\left(-{\epsilon Qx^+\over2}\right).
\ee 
The corresponding effective Newton constant is
\be
G_N^{\rm eff}\sim e^{2\phi}(\ell_s^{\rm eff})^8={\ell_s^{10}\over\epsilon^2R^2}\exp(-2\epsilon Qx^+).
\ee
Comparing the effective string length and Newton constant to the energy scale of interest, we see that in the limit $\epsilon\to 0$
\be
\epsilon E^-\ell_s^{\rm eff}\rightarrow0,\ \ (\epsilon E^-)^8G_N^{\rm eff}\rightarrow0.
\ee
The fact that the effective string length is small implies that massive open strings decouple. The smallness of the effective Newton constant means that closed strings decouple as well. We have thus derived the matrix big bang description \cite{Craps:2005wd}.  
 
To summarize, the DLCQ of type IIA string theory with a lightlike linear dilaton in a sector with $N$ units of lightcone momentum is given by a low-energy limit of the worldvolume theory of $N$ D1-branes in a certain time-dependent background. This is (1+1)-dimensional super-Yang-Mills theory with constant coupling on the Milne orbifold, or equivalently, (1+1)-dimensional super-Yang-Mills theory on a cylinder with a coupling that depends exponentially on time. To get uncompactified type IIA string theory with a lightlike linear dilaton, one eventually has to take a large $N$ limit.  

Now that we have a matrix string description for the lightlike linear dilaton background, let us ask whether it is weakly or strongly coupled. The lightcone energy $\epsilon E^-$ with respect to $x^+$ corresponds to world-sheet energy $E^-\ell_s/R$ with respect to $\tau$ (see \eq{xtau}). If we make the Yang-Mills coupling dimensionless by dividing by the world-sheet energy scale $E^-\ell_s/R$, we find
\be
g_{YM}\left({E^-\ell_s\over R}\right)^{-1}\sim{R\over E^-\ell_s^2}\,\exp\left({Q\tau\ell_s
\over R}\right)\sim{N\over p^+E^-\ell_s^2}\,\exp\left({Q\tau\ell_sp^+\over N}\right).
\ee
Thus for any finite $N$, the theory is weakly coupled at early times and strongly coupled at late times. If $N$ is strictly infinite, the theory is always strongly coupled. The question of how to take the large $N$ limit is important. For now, let us assume it is OK to work with finite $N$.

%%%%%%%%%%%%%%%%%%%%%%%%%%%%%%%%%%%%%%%%%%
\subsection{The one-loop effective potential} 
 
For the matrix theory description of Minkowski spacetime, supersymmetry prevented a potential from being generated for the eigenvalues of the matrices $X^i$. This was crucial for the emergence of space. The lightlike linear dilaton background preserves 16 supersymmetries. However, for any $N>0$ (i.e. for any $p^+>0$) these are all spontaneously broken: the $N$ units of lightcone momentum break the supersymmetries that the linear dilaton background preserved. The worldsheet theory has no unbroken supersymmetry and one expects a potential to be generated quantum mechanically. So does space really emerge at late times in the matrix big bang model? We will see \cite{Craps:2006xq} that for separation $b$ between two eigenvalues and for late times:
\be
\int \sqrt g\, V_{\rm 1-loop}(b)\sim -\int d\tau d\sigma\left(b\over g_s\right)^{1/2}\exp\left(-{Cb\over g_s}\right)
\ee
with $C$ a positive constant. This suggests that the potential turns off at late times ($g_s\to0$), where a spacetime description is expected.

What is this one-loop effective potential $V_{\rm 1-loop}$? A vacuum configuration of the classical theory \eq{matrixstringaction} corresponds to constant matrices $X^i$ that mutually commute. By a gauge transformation, we can make the matrices diagonal. The quartic potential then gives masses to the off-diagonal matrix elements: $X^i_{ab}$ has mass squared
\be
m^2={1\over g_s^2}||X_{aa}-X_{bb}||^2.
\ee
The diagonal matrix elements correspond to flat directions, at least classically. What happens quantum-mechanically, when we integrate out the (massive) off-diagonal modes? Does that generate an effective potential for the (massless) diagonal modes? In cases with enough supersymmetry, this doesn't happen because the contribution from integrating out the bosons is cancelled by the contribution from integrating out the fermions. For the matrix big bang, supersymmetry is broken on the worldsheet of the matrix string, so a potential may be generated. We will compute the one-loop potential, which should be a good approximation at least when $g_{YM}$ is small (close to the big bang).

We now turn to the computation of the effective potential \cite{Craps:2006xq}. One could in principle work with super-Yang-Mills theory on the cylinder and integrate out fields with time-dependent masses. However, the most convenient way to compute the potential is by viewing the matrix big bang action as super-Yang-Mills theory with constant coupling on (the future cone of) the Milne orbifold. The strategy for the computation is as follows. First, the contribution to the one-loop effective action from integrating out a heavy field can be obtained from the propagator of the heavy field in the light-field background of interest. To compute the effective potential, one chooses a constant light-field background. Second, the propagator for a field on (the future cone of) the Milne orbifold can be obtained from a Minkowski space propagator by the method of images. 

First we show how the one-loop effective potential can be obtained from the propagator of the heavy fields. Consider a bosonic field with kinetic operator $K$. For instance, for a (heavy) scalar field in two-dimensional Minkowski space:
\be
K=2\,{\partial^2\over\partial\xi^+\partial\xi^-}+b^2.
\ee
Here $b$ is the mass of the heavy scalar. In the case of interest, it will be the constant expectation value of a light field (the distance between two eigenvalues). Path-integrating over the (heavy) scalar gives a determinant: 
\be
{\rm det}^{-1/2}(K).
\ee
In Feynman diagram language, this determinant is the sum of all vacuum diagrams, connected or disconnected. What appears in the effective action is instead the sum of all one-particle-irreducible diagrams in the light-field background. At one-loop, this corresponds to the logarithm
\be
-i \int V_{\rm 1-loop}={\rm log\ det}^{-1/2}(K)=-{1\over2}{\rm Tr\ log}(K).
\ee
Denote the propagator by $G(\xi,\xi^\prime;b^2)$. Then
\be
e^{tK}(\xi,\xi^\prime)=\oint{dz\over2\pi i} {e^{tz}\over z-K}=-\oint{dz\over2\pi i}e^{tz}G(\xi,\xi^\prime;b^2-z).
\ee
Therefore
\be
-{1\over2}{\rm Tr\ log}(K)={1\over2}\int d^2\xi\int {dt\over t}e^{-it(K-i\epsilon)}(\xi,\xi)
\ee
can be computed from the propagator. Our task is to find the propagator for the various off-diagonal modes of super-Yang-Mills theory in the Milne orbifold, to which we turn next.

We consider super-Yang-Mills theory with constant coupling in the Milne orbifold. The only time-dependence in the problem is in the background metric. Consider a classical vacuum configuration where two eigenvalues are separated by a constant distance $b$. In the Milne description, this leads to off-diagonal modes with constant mass $b$. In lightcone coordinates on the Minkowski covering space $ds^2=-2d\xi^+d\xi^-$, the kinetic operator is
\be
K=2{\partial^2\over\partial\xi^+\partial\xi^-}+b^2.
\ee
The propagator on the Milne orbifold can be easily obtained from the Minkowski propagator by the method of images. The only subtlety is that the action of the discrete boost depends on the spin $s$:
\be
G_{s}(\xi, \xi'; b^{2}) = \sum_{n} \int {d^2p \over (2\pi)^{2}} \, {\exp\left({-i p^{-} (\xi^{+} -e^{2\pi \tilde Q \ell_{s} n} \xi^{+'}) - i p^{+} (\xi^{-} - e^{-2\pi \tilde Q \ell_{s} n} \xi^{-'}) 
+ 2\pi\tilde Q \ell_{s} n s} \right) \over - 2 p^{+} p^{-} + b^{2}},
\ee
with $\tilde Q=Q\ell_s/R$. This propagator gives rise to the kernel
\bea \label{finalker}
e^{-it H_{s}} (\xi, \xi) & =&    \sum_{n} \int {dp^{+}dp^{-} \over (2\pi)^{2}} \,  \exp\Big(-i t\left[ b^{2} - 2 p^{+} p^{-} \right] \\ \nonumber && + \left[-{i p^{-} \xi^{+}( 1- e^{2\pi \tilde Q \ell_{s} n} ) - i p^{+} \xi^{-}(1 - e^{-2\pi \tilde Q \ell_{s} n} ) + 2\pi \tilde Q \ell_{s} n s} \right]  \Big) \\
&= & \nonumber  \sum_{n} {1\over (2\pi) 2t} \exp\Big(- i t b^{2} - i {\xi^{-} \xi^{+} \over 2 t} ( 1- e^{2\pi \tilde Q \ell_{s} n} ) (1 - e^{-2\pi \tilde Q \ell_{s} n}) + 2\pi \tilde Q \ell_{s} n s  \Big)\\
&= & \nonumber  \sum_{n} {1\over (2\pi) 2t} \exp\Big(- i t b^{2} +2 i {\xi^{-} \xi^{+} \over  t} \sinh^2(\pi \tilde Q\ell_s n) + 2\pi \tilde Q \ell_{s} n s  \Big),
\eea
which leads to the following potential term in the action:
\be\label{potentialterm}
\int V_{eff}(b)=i\int d^2\xi\int{dt\over 2t}\sum_{\rm helicities}(-1)^{2s}e^{-it(H_s-i\epsilon)}(\xi,\xi).
\ee
Since ghosts cancel the contributions of two scalars, each supersymmetry multiplet effectively contributes one $s=1$, four $s=1/2$, six $s=0$, four $s=-1/2$ and one $s=-1$ states. Therefore,
\be
\sum_{\rm helicities}(-1)^{2s}e^{2\pi \tilde Q\ell_ss}=\left(e^{\pi \tilde Q\ell_sn/2}-e^{\pi \tilde Q\ell_sn/2}\right)^4=16\sinh^4(\pi \tilde Q\ell_sn/2).
\ee
The potential term \eq{potentialterm} thus reads
\be
\int d^2\xi \sum_{n=-\infty}^\infty\left({2i\over\pi}\right)\sinh^4(\pi\tilde Q\ell_s n/2)
\int_0^\infty {dt\over t^2}\exp\left(-itb^2+{i\over t}2\sinh^2(\pi\tilde Q\ell_s n)\xi^+\xi^-\right).
\ee
Analytically continuing the Schwinger parameter, $t=-it'$, this becomes
\be
-\int d^2\xi \sum_{n=-\infty}^\infty{2\over\pi}\sinh^4(\pi\tilde Q\ell_s n/2)
\int_0^\infty {dt^\prime\over (t^\prime)^2}\exp\left(-t^\prime b^2-{1\over t^\prime}2\sinh^2(\pi\tilde Q\ell_s n)\xi^+\xi^-\right)
\ee
\be\label{potentialBessel}
=-\int d^2\xi \sum_{n=-\infty}^\infty{2\over\pi}{b\sinh^4(\pi\tilde Q\ell_s n/2)\over [2\sinh^2(\pi\tilde Q\ell_sn)\xi^+\xi^-]^{1/2}} \; K_1\!\left(\sqrt{8b^2\sinh^2(\pi\tilde Q\ell_sn)\xi^+\xi^-}\right),
\ee
where $K_1$ is a modified Bessel function, with asymptotic behavior
\bea
K_1(z)\approx{1\over \sqrt z}e^{-z}&&\ \ \ \ (z\gg 1);\label{asymplate}\\
K_1(z)\approx {1\over z}&&\ \ \ \ (z\ll 1).\label{asympearly}
\eea

Now we investigate the behavior of the one-loop effective potential for late and early times.

{}To obtain the very late time behavior, we  use the asymptotic behavior \eq{asymplate} to write \eq{potentialBessel} as
\be\label{asympot}
\int V_{eff}\approx-\int d^2\xi\, {2^{3/4}b^{1/2}\sinh^4(\pi\tilde Q\ell_s/2)\over\pi(\xi^+\xi^-)^{3/4}\sinh^{3/2}|\pi\tilde Q\ell_s|}\,\exp\left(-\sqrt{8b^2\sinh^2(\pi\tilde Q\ell_s)\xi^+\xi^-}\right),
\ee
the dominant contribution coming from $n=\pm1$. This is hugely suppressed at late times. There is an intuitive way to understand this phenomenon. The extent to which supersymmetry is broken is controlled by the size of the Milne circle (parametrized by $\sigma$). At $\tau \to \infty$, the circle becomes large and supersymmetry is effectively restored. While there was no a priori reason for us to see the potential vanish at late times at just $1$-loop, it is a result in perfect agreement with the claim that at late times, this theory flows to string field theory in the lightlike linear dilaton background~\cite{Craps:2005wd}. In such a theory, there is no perturbative potential. However, if we express~\eq{asympot}\ in terms of the perturbative string coupling, 
\be\label{pertpot}
\int \sqrt g\, V_{\rm 1-loop}(b)\sim -\int d\tau d\sigma\left(b\over g_s\right)^{1/2}\exp\left(-{Cb\over g_s}\right)
\ee
with $C$ a positive constant, we see that there is a non-perturbative potential that appears to be generated by D-branes. If higher loop corrections to the potential are more suppressed then we might hope to compute this potential directly in string theory. 

Now let us discuss the early time behavior of the potential. We have seen that the summand in \eq{potentialBessel} decreases quickly as a function of $n$ when the argument of the modified Bessel function is larger than one. However, at early times, $b^2\xi^+\xi^-\ll1$, the argument is smaller than one for a range of values of $n$, so we should use the asymptotic behavior \eq{asympearly}. We find
\bea
\int V_{eff}&\approx&-\int d^2\xi \sum_{n}{2\over\pi}{b\sinh^4(\pi\tilde Q\ell_s n/2)\over [2\sinh^2(\pi\tilde Q\ell_sn)\xi^+\xi^-]^{1/2}} \;
{1\over\sqrt{8b^2\sinh^2(\pi\tilde Q\ell_sn)\xi^+\xi^-}}\\ \nonumber
&=&-\int d^2\xi\, {1\over8\pi\xi^+\xi^-}\sum_n\tanh^2(\pi\tilde Q\ell_sn/2)\approx \int d^2\xi\, {1\over8\pi^2\tilde Q\ell_s\xi^+\xi^-}\log(2b^2\xi^+\xi^-),
\eea
where the sum was taken over those values of $n$ for which the argument of the modified Bessel function is smaller than 1. At early times, we thus find an attractive one-loop potential between two eigenvalues.

In \cite{Li:2005ai}, the one-loop effective potential was claimed to vanish. However, it seems that the authors of \cite{Li:2005ai} compute a time-averaged version of the potential rather than the time-dependent potential itself, which might help explain the apparent contradiction with our result.

To summarize, off-diagonal modes (``stretched open strings'') have masses proportional to differences between expectation values of diagonal modes (``distances between D1-branes''). Integrating out the off-diagonal modes induces an effective potential for the diagonal modes. We computed this potential from the propagator of the off-diagonal modes, which is easily obtained from the method of images (since the worldsheet is an orbifold of flat spacetime). The one-loop effective potential is attractive and turns off at late times, i.e.\ the regime where a spacetime description is expected.

The matrix big bang model has been generalized in various directions. More general supergravity solutions for which a matrix theory description can be given are discussed in \cite{Li:2005sz, Li:2005ti, Das:2005vd, Chen:2005mg, Chen:2005bk, Ishino:2005ru, Das:2006dr, Chen:2006rm, Ishino:2006nx}. A similar matrix theory description of the ``null-brane'', a resolved version of the parabolic orbifold we mentioned before \cite{Figueroa-O'Farrill:2001nx, Simon:2002ma, Liu:2002kb, Fabinger:2002kr}, was given in \cite{Robbins:2005ua, Martinec:2006ak}. Recently, similar lightlike singularities have also been studied in an AdS/CFT context \cite{Chu:2006pa, Das:2006dz, Lin:2006ie}. D-branes in the lightlike linear dilaton background have been studied in \cite{Nayak:2006dm}.  

From these lectures, we can draw the following conclusions. General relativity breaks down near spacetime singularities. String theory has been very successful in resolving static singularities. Lightlike and spacelike singularities have proven to be hard to resolve in perturbative string theory. The matrix big bang is a proposal for a non-perturbative description of a lightlike singularity. Near the big bang, the model describes weakly coupled matrices. At late times, spacetime emerges dynamically. A potential problem for the emergence of spacetime, related to the absence of unbroken supersymmetry, appears to be harmless in this model. 

We end with a number of open questions regarding the matrix big bang model:
\begin{enumerate}
\item
Do higher-loop contributions to the effective potential also turn off at late times, so that the flat directions and the spacetime interpretation are (approximately) preserved?
\item
What are the derivative terms in the effective action?
\item
Can time be extended beyond the Big Bang? One idea is to extend the Milne worldsheet to the past cone (and the ``whisker'' regions), but then one has to confront the propagation of field theory degrees of freedom through the Milne singularity. Some preliminary work is in \cite{Craps:2005wd} and \cite{Hikida:2005ec}. 
\item
What are the string theory observables in a spacetime with a Big Bang?
\item
One could choose an initial state in the weakly coupled SYM near the Big Bang. Are there ways to compute what this state evolves to at late times? The challenge is that the intermediate time regime is at intermediate coupling.
\item
How should one take the large $N$ limit? Is it OK to keep $N$ finite in the computations and take it to infinity at the very end?
\end{enumerate}
 
%%%%%%%%%%%%%%%%%%%%%%%%%%%%%%%%%%%%%%%%%%%%%%%%%%%%%%%%%%%%%%%%%%%%%%%%%%%%%%%%%%%%%%%%%%%%%%%%%%%%%%%%%%%%%%%%%%%%%%%%%%%%%%%%%%%%%%%%%%%%%%%%%%%%%%%%%%%%%%%%%%%%%%%%%%%%%%%%%%%%%%%%%%%%%%%%%%%%%%
\section*{Acknowledgments}
I would like to thank the organizers for a very pleasant and stimulating school and for the opportunity to present these lectures. My own contributions to the material discussed in these lectures were obtained in collaborations involving M.~Berkooz, M.~Bill\'o, D.~Kutasov, B.~Ovrut, A.~Rajaraman, G.~Rajesh, F.~Roose, S.~Sethi and E.~Verlinde; I wish to thank all of them for many useful and enjoyable discussions. This work is supported in part by the Belgian Federal Science Policy Office through the Interuniversity Attraction Pole P5/27, by the European Commission FP6 RTN programme MRTN-CT-2004-005104 and by the ``FWO-Vlaanderen'' through project G.0428.06.

%%%%%%%%%%%%%%%%%%%%%%%%%%%%%%%%%%%%%%%%%%%%%%%%%%%%%%%%%%%%%%%%%%%%%%%%%%%%%%%%%%%%

%%%%%%%%%%%%%%%%%%%%%%%%%%%%%%%%%%%%%%%%%%%%%%%%%%%%%%%%%%%%%%%%%%%
\end{document}